\documentclass[longauth]{aa}
\usepackage{graphicx}
\usepackage{txfonts}
\usepackage{graphicx}   
\usepackage{multirow}
\usepackage{ulem}
\usepackage{subfig}
\usepackage{upgreek}
\usepackage{hyperref}
\usepackage{placeins}
\hypersetup{
    citecolor=blue,
    colorlinks=true,
    linkcolor=blue,
    filecolor=magenta,      
    urlcolor=blue
    }

\graphicspath{{Figures/}}
\begin{document} 

   \title{Metal content of the circumgalactic medium around star-forming galaxies at z $\sim$ 2.6 as revealed by the VIMOS Ultra-Deep Survey}
   \subtitle{}
   \author{M\'endez-Hern\'andez, H.,
          \inst{1}
          Cassata, P.\inst{2,3}, 
          Ibar, E.\inst{1},
          Amor\'in, R. \inst{4,5},
          Aravena, M. \inst{6},
          Bardelli, S. \inst{7},
          Cucciati, O. \inst{7},
          Garilli, B. \inst{8},
          Giavalisco, M. \inst{9},
          Guaita, L. \inst{10},
          Hathi, N. \inst{11},
          Koekemoer, A. \inst{11},
          Le Brun, V. \inst{12},
          Lemaux, B.C. \inst{13,14},
          Maccagni, D. \inst{8},
          Ribeiro, B. \inst{12},
          Tasca, L. \inst{12},
          Tejos, N. \inst{16},
          Thomas, R. \inst{17},
          Tresse, L. \inst{12},          
          Vergani, D. \inst{7},
          Zamorani, G. \inst{7},
          Zucca, E.  \inst{7}
          }
   \institute{Insituto de F\'isica y Astronom\'ia, Universidad de Valpara\'iso,
              Avda. Gran  Breta\~na 1111, 2340000 Valpara\'iso, Chile\\
              \email{hugo.mendez@postgrado.uv.cl}
         \and
             Dipartimento di Fisica e Astronomia Galileo Galilei, Universit\`a 
             degli Studi di Padova Vicolo del $L'$Osservatorio 3, 35122 Padova Italy
        \and
             INAF Osservatorio Astronomico di Padova, vicolo dell'Osservatorio 5, I-35122 Padova,Italy
        \and
            Instituto de Investigaci\'on Multidisciplinar en Ciencia y Tecnolog\'ia, Universidad de La Serena, Ra\'ul Bitr\'an, 1305 La Serena, Chile
        \and
            Departamento de Astronom\'ia, Universidad de La Serena, Av. Juan Cisternas 1200 Norte, La Serena, Chile             
        \and
            N\'ucleo de Astronom\'ia, Facultad de Ingenier\'ia y
            Ciencias, Universidad Diego Portales, Av. Ej\'ercito 441, Santiago, Chile
        \and 
            INAF-Osservatorio di Astrofisica e Scienza dello Spazio di Bologna, 
            via Gobetti 93/3, 40129 Bologna Italy
        \and
            INAF-IASF, via Bassini 15, I-20133, Milano, Italy
        \and
            Department of Astronomy, University of Massachusetts Amherst, 710 N. Pleasant St, Amherst, MA 01003, USA
        \and
            Departamento de Ciencias Fisicas, Facultad de Ciencias Exactas, 
            Universidad Andres Bello, Fernandez Concha 700, Las Condes, Santiago, Chile
        \and
            Space Telescope Science Institute, 3700 San Martin Drive, Baltimore, MD 21218, USA
        \and
            Aix Marseille Univ, CNRS, CNES, LAM, Marseille, France
        \and
            Gemini Observatory, NSF's NOIRLab, 670 N. A'ohoku Place, Hilo, Hawai'i, 96720, USA
        \and
            Department of Physics and Astronomy, University of California, Davis, One Shields Ave., Davis, CA 95616, USA
        \and
            Space Telescope Science Institute, 3700 San Martin Drive, Baltimore, MD 21218, USA
        \and
            Instituto de F\'isica, Pontificia Universidad Cat\'olica de Valpara\'iso, Casilla 4059, Valpara\'iso,
            Chile
        \and
            European Southern Observatory, Av. Alonso de C\'ordova 3107, Vitacura, Santiago, Chile
            }
   \date{Received XXXX XX, 2021; accepted XXXX XX, 2022}
  \abstract
   {The circumgalactic medium (CGM) is the location where the interplay
   between large-scale outflows and accretion onto galaxies occurs. Metals in
   different ionization states flowing between the circumgalactic and
   intergalactic mediums are affected by large galactic outflows and 
   low-ionization state inflowing gas. Observational studies on their spatial
   distribution and their relation with galaxy properties may provide
   important constraints on models of galaxy formation and evolution.}
   {The main goal of this paper is to provide new insights into the spatial
   distribution of the circumgalactic of star-forming galaxies at $1.5 < z <
   4.5$ ($\langle z\rangle\sim$2.6) in the peak epoch of cosmic
   star formation activity in the Universe. We also look for possible
   correlations between the strength of the low- and high-ionization
   absorption features (LIS and HIS) and stellar mass, star formation rate,
   effective radius, and azimuthal angle $\phi$ that defines the location of
   the absorbing gas relative to the galaxy disc plane.}
   {The CGM has been primarily detected via the absorption features that it
   produces on the continuum spectrum of bright  \textit{background} sources.
   We selected a sample of 238 close pairs from the VIMOS Ultra Deep Survey to
   examine the spatial distribution of the gas located around star-forming
   galaxies and generate composite spectra by co-adding spectra of \textit
   {background} galaxies that provide different sight-lines across the CGM of
   star-forming galaxies.}
   {We detect LIS (C\,{\sc ii} and Si\,{\sc ii}) and HIS (Si\,{\sc iv}, C\,
   {\sc iv}) up to separations $\langle b \rangle=$\,172\,kpc and 146\,kpc.
   Beyond this separation, we do not detect any significant
   signal of CGM absorption in the \textit{background} composite spectra. Our
   Ly$\upalpha$, LIS, and HIS rest-frame equivalent width ($W_{0}$) radial
   profiles are at the upper envelope of the $W_{0}$ measurements at lower
   redshifts, suggesting a potential redshift evolution for the CGM gas
   content producing these absorptions. We find a correlation between C\,
   {\sc ii} and C\,{\sc iv} with star formation rate and  stellar mass, as well as
   trends with galaxy size estimated by the effective radius and azimuthal
   angle. Galaxies with high star formation rate (log[SFR/(M$_{\odot}$yr$^
   {-1})]>1.5$) and stellar mass (log[M$_{\bigstar}$/M$_{\odot}$]>10.2) show
   stronger C\,{\sc iv} absorptions compared with those low SFR (log[SFR/(M$_
   {\odot}$yr$^{-1})]<0.9$) and low stellar mass (log[M$_{\bigstar}$/M$_
   {\odot}$]<9.26). The latter population instead shows stronger  C\,
   {\sc ii} absorption than their more massive or more star-forming
   counterparts. We compute the C\,{\sc ii} / C\,{\sc iv} $W_{0}$ line ratio
   that confirms the C\,{\sc ii} and C\,{\sc iv} correlations with impact
   parameter, stellar mass, and star formation rate. We do not find any
   correlation with $\phi$ in agreement with other high-redshift studies and
   in contradiction to what is observed at low redshift where  large-scale
   outflows along the minor axis forming bipolar outflows are detected.}
   {We find that the stronger C\,{\sc iv} line absorptions in the outer
   regions of these {star-forming} galaxies could be explained by stronger
   outflows in galaxies with higher star formation rates and stellar masses that are 
   capable of projecting the ionized gas up to large distances and/or by
   stronger UV ionizing radiation in these galaxies that is  able to ionize the gas
   even at large distances. On the other hand, low-mass galaxies show
   stronger C\,{\sc ii} absorptions, suggesting larger reservoirs of cold gas
   that could be explained by a softer radiation field unable to ionize
   high-ionization state lines or by the galactic fountain scenario where
   metal-rich gas ejected from previous star formation episodes falls back to
   the galaxy. These large reservoirs of cold neutral gas around low-mass
   galaxies could be funnelled into the galaxies and eventually provide the
   necessary fuel to sustain star formation activity.}

   \titlerunning{The CGM metal content around star-forming galaxies at $z\sim$2.6 revealed by VUDS}
   \authorrunning{M\'endez-Hern\'andez, H.}
   \maketitle
%
\section{Introduction}
\noindent
The circumgalactic medium (CGM, 10kpc-$\sim$300 kpc) is the gas reservoir
between the interstellar medium (ISM, {$\lesssim10$kpc}) and the
intergalactic medium (IGM, $\gtrsim300$kpc). It is characterized as an active
interface where galaxies reprocess their baryonic material. Up to $\sim$50\%
of the total baryonic mass is found in the CGM at low redshift
($z\sim0.2$) \citep{Wolfe05,Werk14,Zheng15} and at high redshift \citep
{Hafen19}, thus representing a significant gas reservoir that can, for
example, feed the ISM with gas to form new stars \citep
{Zhu13a,Thom12,Richter16}. Supernovae explosions or strong stellar winds can
deposit metals in the surrounding medium, gas that is mixed with pristine gas
accreted from the IGM (\citealt
{Prochaska09,Bauermeister10,Tumlinson11a,Tumlinson17,Kacprzak17}).

Since the first discoveries of circumgalactic gas around star-forming
galaxies \citep{Boksenberg78,Kunth84,Bergeron86}, its study by direct
detection of emission has been a challenge \citep
{vandeVoort13,Burchett21}. The CGM has been primarily detected via the
absorption features that it produces on light from \textit
{background} sources. In a pioneering study at high redshift
($z\gtrsim 1.5$), 
\cite{Adelberger03,Adelberger05a} studied the dependence of the absorption
strength on the projected angular separation between \textit
{foreground} and \textit{background} galaxies  (i.e. impact parameter $b$);
they reported absorption detection at impact parameters up to 40\,kpc and
found that absorption strength weakens significantly at larger separations.
Thanks to their brightness, quasars are typically used as \textit
{background} sources to probe the CGM around \textit{foreground} galaxies
\citep{Bergeron91,Steidel94,Kacprzak10}, which allows  the signal produced by
extremely low column densities (N$_{\rm{HI}}\simeq10^{12}$ cm$^{-2}$) to be
identified independently of the properties of the targeted galaxy
(e.g.  luminosity and/or redshift) 
\citep{Tumlinson17,Peroux20a}. This technique has been used successfully to
probe the CGM of galaxies up to  redshifts $z\sim$5 \citep{Matejek12} and
has led to the discovery that the CGM is gas rich and has a multi-phase
nature
\citep
{Adelberger03,Steidel94,Steidel95,Songaila01,Tripp98,Kacprzak10,Werk16}. 

The CGM has been also characterized using the `down-the-barrel technique',
which uses the targeted galaxy itself as a \textit{background} source.
Absorption lines redshifted with respect to the galaxy rest-frame give
evidence of the presence of gas flowing towards the galaxy. This result has
been interpreted as a strong indication of inflowing material onto the host
galaxy \citep{Sato09,Rubin12,Martin12}, while outflows have been identified
as blueshifted absorptions in galaxy spectra \citep
{Martin05,Chen10}. Although the location of the gas producing the detected
absorption is unconstrained, this technique has been successful in studying
galactic inflows and outflows from the spectroscopy of star-forming galaxies
up to $z\sim2-3$ \citep{Kornei12,Rudie12a,Heckman15,Heckman16}. 

Recently, three more techniques have been used to explore the CGM. The first,
gravitational-arc tomography, takes advantage of strong gravitational lensing
and uses giant bright lensed arcs as
\textit{background} sources to map the CGM of \textit{foreground} galaxies
providing a tomographic view of the absorbing gas \citep
{Lopez18,Claeyssens19,Lopez20,Tejos21,Mortensen21}. The second technique
takes advantage of deep three-dimensional datacubes observations to study
the cold CGM of high-redshift ($z > 2$) star-forming galaxies, and has
reported the ubiquitous presence of Ly$\upalpha$ haloes in these galaxies 
\citep{Steidel11,Matsuda12,Momose14,Wisotzki16,Leclercq17,Chen20}, whose line
properties are correlated to their spatial location 
\citep{Leclercq20} and that can extend up to 4Mpc beyond the CGM 
\citep{Chen20,Bacon21}. The third technique uses the spectra of bright
afterglows of long gamma-ray bursts (GRBs) to derive the kinematic
properties of the CGM gas around their host galaxies and constrain the
physical properties \citep{Gatkine19,Gatkine22}.

Alongside these diverse techniques used to study the CGM, the large
spectroscopic extragalactic surveys have allowed  the statistical extraction
and analysis of the weak signals from different metal absorption using the
stacking of hundreds of spectra \citep{Steidel10}. These large datasets have
been helpful  to overcome the limitations of finding galaxy--QSOs pairs at
concordant redshifts \citep{Steidel94,Bouche07}, increasing statistics, and
facilitating vast parameter space exploration 
\citep{York06,Bordoloi11,Zhu13a}. At low redshift ($z_{\rm{med}}\sim$0.5) CGM
analyses have been focused on the study of absorption lines of
Ly$\upalpha$ \citep{Chen01b}, C\,{\sc iv} \citep{Chen01a}, O\,
{\sc vi} \citep{Tumlinson11a}, and Mg\,{\sc ii} \citep
{Bowen95,Bouche07,Steidel94}, and their dependence on stellar mass \citep
{Bordoloi11}, inclination \citep{Kacprzak10}, and azimuthal angle \citep
{Shen12,Bordoloi14a}. At higher redshifts, the observed metal absorption
lines are mostly limited to Si\,{\sc ii}, C\,{\sc ii}, C\,{\sc iv}, and
Si\, {\sc iv};   the first two are called low-ionization state (LIS, T=$10^
{4-4.5}$K) lines, and  the last two high-ionization state (HIS, T=$10^
{4.5-5.5}$K) lines \citep{Steidel10}. At high redshift (z>2) several authors
have reported the presence of high-velocity outflows. \cite
{Lehner14} demonstrated that O\,{\sc vi} successfully probes outflows in
star-forming galaxies, and reported velocity widths in the range
200-400kms$^{-1}$. Next, \cite{Du16} analysed several ionization lines
(e.g. Si\,{\sc ii}, Fe\,{\sc ii}, Al\,{\sc ii}, Ni\,{\sc ii}, Al\,{\sc iii},
C\,{\sc iv}) to probe the multi-phase nature of the CGM and detected C\,
{\sc iv} blueshifted offsets concordant  with 76kms$^{-1}$ velocity
outflows. Their results show a direct link between C\,{\sc iv} absorption
and star formation rate. Later on, \cite{Jones18} reported velocity outflows
of $\sim$150km$^{-1}$, as shown by several LIS and HIS absorptions in nine
gravitationally lensed star-forming galaxies($z\simeq2-3$), suggesting that
galaxy outflows regulate the galaxy chemical evolution. Similar outflow
detections inferred from different ionization absorption lines have been
reported, and  stellar mass and star formation rate have been invoked as
their main drivers \citep
{Zhu13a,Turner14,Trainor15,Gatkine19,Price20}. However, as suggested
by \cite{Dutta21}, LIS and HIS absorption detections (from which velocity
outflows can be inferred), could be affected by large-scale environmental
processes \citep{Dutta21,WangS22} or their available neutral gas
content \citep{Berry12,Oyarzun16,Du18}, and their correlation with stellar
mass and star formation rate might  in fact be a consequence of a
main-sequence offset rather than simply correlated with the star formation
rate or stellar mass \citep{Cicone16,Gatkine22}.

Numerical simulations consider that the CGM comes from gas accreted from the
IGM, followed by stellar winds from the central galaxy and the gas ejected or
stripped from satellites \citep{Hafen19,Christensen18}. Simulations reveal
that the accretion efficiency depends on the galaxy stellar mass, decreasing
from $\sim$80\%\  for M$_{\bigstar} \sim 10^{6}$M$_{\odot}$ galaxies to
$\sim$60\%\   for M$_{\bigstar} \sim 10^{10}$M$_{\odot}$ galaxies
\citep{Dekel05,Keres05,Hafen20}. Once accreted, this material can remain in
the CGM for billions of years leading to a well-mixed halo gas before it
interacts with the ejected large-scale stellar winds produced by starbursts
\citep{Martin05,Weiner09}. At $z=2$ most metals are found to be located in the
ISM  or stars of the central galaxy, and by $z=0.25$ most of it  will end up
in the CGM and IGM (e.g. \citealt
{Peeples14,Angles17,Hafen20,Oppenheimer16,Nelson21}).

The way in which this pristine material is accreted into galaxies may depend
on its location relative to the galaxy disc plane defined by the azimuthal
angle ($\phi$). Various observational studies highlight a correlation between
the strength of the Mg\,{\sc ii} LIS absorption line and the azimuthal
angle 
\citep{Bordoloi11,Bordoloi14a,Kacprzak11b,Bouche12,Bouche13}. Although weak
Mg\,{\sc ii} detections along the minor axis have been reported
(e.g. \citealt{Lan14}), it has been found that strong Mg\,{\sc ii}
absorptions are preferably detected along the major-axis of galaxies,
suggesting the presence of inflowing material potentially feeding future
star formation. On the other hand, C\,{\sc iv} and O\,{\sc vi} (HIS lines)
absorptions seem stronger along the minor axis, probably evidence of strong
stellar winds enriching the CGM 
\citep{Tumlinson11a,Kacprzak15a}. Recently, 
\cite{Peroux20b} used cosmological hydrodynamical simulations to examine the
physical properties of the gas located in the CGM of star-forming galaxies
($z<1$) as a function of angular orientation. They found that the CGM
properties vary strongly with the impact parameter, stellar mass, and
redshift. They reported a higher average CGM metallicity at large impact
parameters ($b>$100kpc) along the minor versus major axes. Moreover, they
tentatively found that the average metallicity of the CGM depends on the
azimuthal angle, showing that the low-metallicity gas preferably inflows
along the galaxy major axis, while outflows are commonly located along the
minor axis, in agreement with previous observations 
\citep{Bordoloi11,Bouche12,Kacprzak12a,Kacprzak15a}. These results present a
picture where star-forming galaxies accrete co-planar gas within narrow
stream-flows providing fresh fuel for the new generation of stars; later,
this population will produce metal-enriched galactic-scale outflows along
the minor axis \citep{Kacprzak17}. However, even though the presence of
bipolar outflows collimated along the minor axis is expected to evolve with
redshift (expected to be ubiquitous at z = 1), it has been difficult to
demonstrate its presence at z = 2, mostly as the result of the absence of
gaseous galactic discs sculpting the outflows \citep{Nelson19}.

In order to better understand the mechanisms of galaxy growth in earlier
phases of the history of the Universe, and to investigate a possible cosmic
evolution of such mechanisms, similar studies at higher redshifts are needed.
Observations along different lines of sight (l.o.s.) can probe different
parts of the CGM, and the identification of low- and high-ionization metal
absorption lines can give information on the possible multi-phase nature of
the CGM. However, using QSO--galaxy pairs to study the CGM is not
straightforward as the identification of the galaxies responsible for the
metal absorptions detected on the spectra of \textit{background} quasars is
not an easy task. Stacking analyses of \textit{background} galaxy spectra
that are located in the vicinity of \textit{foreground} galaxies provides an
alternative tool to overcome sensitivity limitations, exploiting large
extra-galactic surveys containing a large number of individual spectra. In
this work we seek to characterize the presence of low- (LIS: \ O\,
{\sc i}+Si\,{\sc ii}, C\,{\sc ii}, Si\,{\sc ii}, Fe\,{\sc ii}),
intermediate- (IIS: Al\,{\sc iii}), and high-ionization (HIS: Si\,{\sc iv},
C\,{\sc iv}) state metal absorption (see Table \ref{Tbl:Lines}) in the CGM of
a star-forming galaxy population at $\langle z \rangle$ $\sim$ 2.6 using UV
spectra obtained from the large VIMOS Ultra Deep Survey (VUDS; \citealt
{LeFevre15}). We use thousands of galaxies with the most reliable redshift
measurements and the highest S/N spectra (with reliability flags 3 and 4;
see \citealt{LeFevre15,Tasca17}), a sample broadly representative of the bulk
of the star-forming galaxy population at these redshifts \citep{Lemaux22}. To
detect the dim signal coming from low- and high-ionization line absorptions
produced in the CGM of these star-forming galaxies, we stack the spectra of
close (in projection) \textit{background} galaxies to establish the metal
distribution in the CGM and explore their dependence on the physical and
morphological properties of star-forming galaxies (e.g.  impact parameter
$b$, star formation rate, stellar mass, galaxy effective radius, and
azimuthal angle) at the peak epoch of cosmic star formation activity in the
Universe. 

The manuscript is organized as follows. Section~\ref{sec:Data} summarizes the
VUDS survey properties relevant to our analyses. Section~\ref
{sec:Ana} presents our stacking  analysis method for measuring the metal
equivalent widths ($W_{0}$), while Section~\ref{sec:Res} presents our $W_
{0}$ results for Ly$\upalpha$, and multiple LIS, IIS, and HIS metal lines
observed our star-forming galaxy sample across different physical and
morphological properties. We discuss our results in Section~\ref
{sec:Dis}, and finally we present our conclusions in Section~\ref
{sec:Con}. Throughout the text we use  $\Lambda$CDM cosmology with H$_
{0}$\,=\,70\,km\,s$^{-1}$\,Mpc$^{-1}$, $\Omega_{\rm M}$=0.3, and $\Omega_
{\Lambda}$=0.7, and distances are given in physical units (kpc).

\section{The VUDS parent sample}\label{sec:Data}

The VUDS has obtained spectra of 5590 galaxies in the redshift range $1.5 < z <
4.5$. A detailed description  of the survey observations,
the methods applied to process the data, and the derived parameters including
the spectroscopic redshifts $z_{\rm{spec}}$ is given in \cite{LeFevre15}. A
description of the VUDS-DR1 first data release can be found in \cite
{Tasca17}. The VUDS spectroscopic targets are selected based on their
photometric  redshifts and observed optical flux; the targets have $z_
{phot}$ + 1$\sigma_{\rm z} \geq$ 2.4 and  i$_{AB} \leq$ 25. The  wavelength
range of each spectrum is between 3600\,$<\uplambda/\AA<$\,9350, accumulating
14\,hrs of integration time in each of the LRBLUE and LRRED grisms of the
VIMOS spectrograph on the ESO Very Large Telescope \citep{LeFevre03}, with a
spectral resolution R\,$\sim$\,230($\sim 7\,\AA$) and reaching a S/N=5 on the
continuum at 8500\,$\AA$. We note that the VUDS observations were taken using
the VIMOS low-resolution multi-slit mode with a minimum slit length optimized
to 6\,arcsec, maximizing the number of observed slits per mask (see \citealt
{Bottini05}).

Standard data processing was performed using the VIPGI environment  
\citep{Scodeggio05}, followed by redshift measurements using the {\sc EZ}
package \citep{Garilli10}. The final UV rest-frame flux-limited sample is
broadly representative of the bulk of the star-forming galaxy population at
these redshifts ($2 < z < 5$) \citep{Lemaux22}. A critical aspect of VUDS is
the large comoving volume covered, totalling   1\,deg$^{2}$ in three
fields: COSMOS \citep{Scoville07}, ECDFS \citep{Giacconi02}, and
VVDS02h \citep{LeFevre05b,LeFevre13b}.

For the purposes of this study the instrumental set-up translates into the
ability to follow lines redder than Ly$\upalpha$ ($\geq1215.6\AA$)  at
redshifts higher than 1.5. We decided to restrict the VUDS sample to 2100
galaxies, selected in the redshift range $1.5 < z < 4.5$ ($z_{\rm
{flag}}=3, 4$, i.e.\ $95-100\%$ probability  of being correct). We note that
the velocity accuracy of redshift measurements is expected to be in the range
d$z$ / (1 + $z$) = 0.0005-0.0007, or an absolute velocity ($\sigma^{-1}_
{v}$) accuracy 150-200 km  \citep{LeFevre13b,LeFevre15,Tasca17}. The rich
broad-band imaging available for galaxies in VUDS is used for spectral energy
distribution (SED) fitting \citep{Tasca15,Thomas17} using the Galaxy
Observed-Simulated SED Interactive Program ({\sc GOSSIP};
\citealt{Franzetti08}) to derive various global galaxy properties, for example
stellar mass, star formation rate (SFR), dust extinction E(B-V), age, 
and metallicity.

\begin{figure}
\centering
\includegraphics[width=8.5cm]{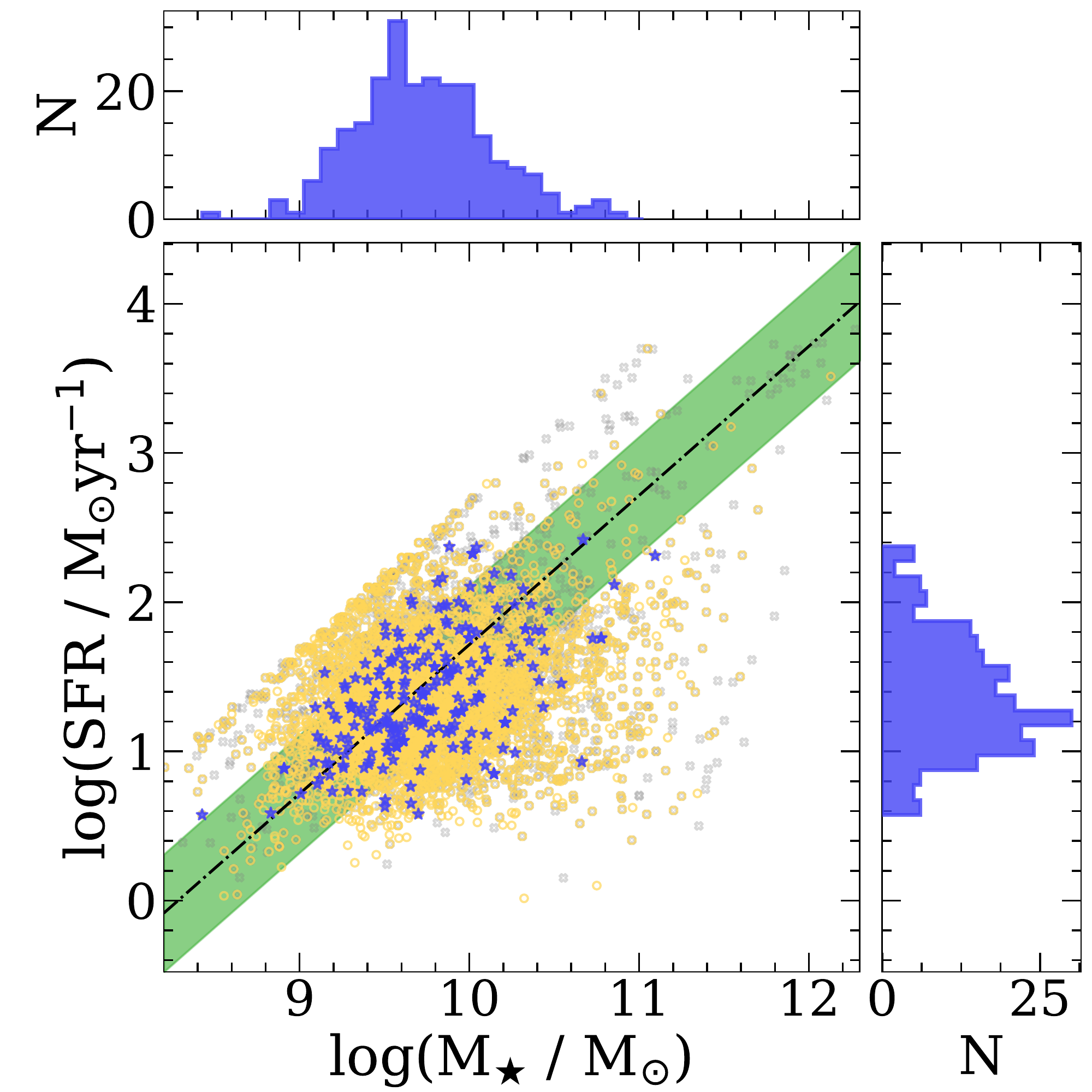}
\caption{Star formation rate versus stellar mass for 5590 star-forming galaxies selected from
the  VUDS survey  at $z>1.5 $ (grey open squares) and  at  $2 < z < 3.5$
(yellow crosses), and the 238 star-forming galaxies (blue stars) considered in
this work. The black dot-dashed line shows the main sequence of star-forming
galaxies   for $1.5 < z < 2.5$ including its $\pm.3$dex scatter
represented by the green shaded region, as defined by \citealt{Hathi16}. The
top and right panels report the distributions in stellar mass and SFR
for the 238 star-forming galaxies evaluated in this work.}
\label{Fig:SFR-STM}
\end{figure}

{\sc GOSSIP} is a tool that performs SED fitting by using a combination of
spectro-photometric measurements from different bands to match a set of
synthetic galaxy spectra based on emission from stellar populations.
{\sc GOSSIP} uses model spectra from galaxy population synthesis models 
\citep{Bruzual03,Maraston05} and uses the probability distribution function
(PDF) of each galaxy parameter to determine the best SED fit. Figure \ref
{Fig:SFR-STM} shows SFR versus stellar mass for 5590 star-forming galaxies
at $z > 1.5$ and  $2 < z < 3.5$
selected from the VUDS survey, and the 238 star-forming galaxies  that we
finally selected and scrutinized in this work (see Section~\ref
{sec:Ana}). These 238 galaxies have stellar masses and star formation rates
of log[M$_{\bigstar}$/M$_{\odot}$])=$9.73\pm0.4$ and log[SFR/(M$_
{\odot}$yr$^{-1}$)]=$1.38\pm0.39$.

\begin{figure*}
\centering
\includegraphics[width=17cm]{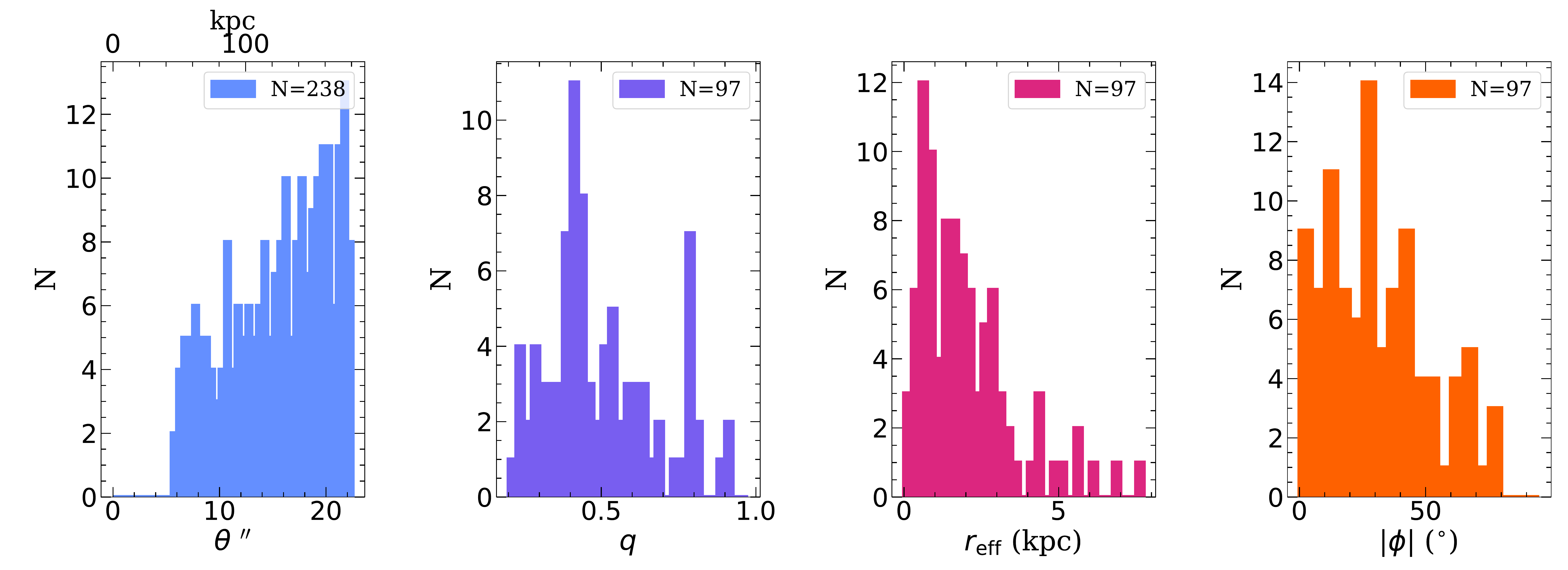}
\caption{Projected angular separation ($\theta$) amongst galaxy 
\textit{foreground-background} galaxy pairs, axial ratio (q), 
effective radius (r$_{\rm{eff}}$), and azimuthal angle ($\phi$) distributions
of the star-forming (\textit{foreground}) galaxies used in our analyses. 
The morphological parameters were obtained from 
the parametric measures of \cite{Ribeiro16}, 
see section \ref{sec:Data} for details.}
\label{Fig:PairsHisto}
\end{figure*}

Exploiting the Hubble Space Telescope (HST) $F814W$ images \citep
{Koekemoer07} that are available as part of the COSMOS Survey \citep
{Scoville07}, morphological parameters have been estimated for 1242
star-forming galaxies ($z_{\rm{flag}}=4, 3, 2, 9$, i.e.\ $75\%$ probability of
being correct) with 9.5<log[M$_{\bigstar}$/M$_{\odot}$]<11.5. In
particular, \cite{Ribeiro16} run {\sc Galfit} \citep{Peng02,Peng10}, a
standard parametric profile-fitting tool, estimating Sersic indices
($n$), azimuthal angles, major-to-minor axis ratios ($q$), and effective
radii ($r_{\rm eff}$). Figure~\ref{Fig:PairsHisto} shows the distribution of
the 97 selected galaxy pairs (see Section~\ref{sec:Ana}) with available
morphological parameters. Table \ref{Tbl:Samples} shows the detailed number
of subsets of galaxies.

\begin{table}
\centering
\caption{\label{Tbl:Samples}
VUDS \textit{fg-bg} galaxy pair sample.}
\begin{tabular}{cc}
  \\
  \hline\hline
  \multicolumn{1}{c}{Sample} &
  \multicolumn{1}{c}{N} \\
  \hline\hline  
Galaxies $z > 1.5$, $z_{\rm{flag}}$=3, 4, 2, 9 \tablefootmark{a}  & 5590 \\ 
Galaxies $z > 1.5$, $z_{\rm{flag}}$=3, 4 \tablefootmark{b} & 2100 \\
\textit{fg-bg} galaxy pairs, $z_{\rm{flag}}$=3, 4 \tablefootmark{c} & 238 \\
Galaxies fitted with {\sc Galfit}, $z_{\rm{flag}}$=3, 4, 2, 9  \tablefootmark{d} & 1242 \\
\textit{fg-bg} galaxy pairs, $z_{\rm{flag}}$=3, 4 + {\sc Galfit}  \tablefootmark{e} & 97 \\
\hline
\end{tabular}
\tablefoot{
Number of objects of our \textit{fg-bg} galaxy pair sample sample
(see Sect. \ref{sec:Data} for details).\\
\tablefoottext{a}{VUDS parent sample.}
\tablefoottext{b}{Star-forming galaxies at $z>1.5$.}
\tablefoottext{c}{Star-forming galaxies with morphological parameters available.}
\tablefoottext{d}{Close \textit{fg-bg} galaxy pairs.}
\tablefoottext{e}{Close \textit{fg-bg} galaxy pairs with morphological 
parameters.}}
\end{table}


\section{Analysis}\label{sec:Ana}

In this work we describe our  probe of  the CGM around galaxies at  $z \sim$ 2.6, near
the epoch at which the cosmic SFR density of the Universe
reaches its peak, by using the spectra of \textit{background (bg)} galaxies
as `spotlights' illuminating the CGM. At these redshifts individual \textit
{bg} galaxies are not bright enough to individually detect low- and
high-ionization state metal absorption lines. Thus, we stack spectra
of \textit{bg} galaxies to look for absorption lines produced by the CGM
around \textit{foreground (fg)} star-forming galaxies. To select a sample of
close galaxy pairs(\textit{fg-bg}) out of the VUDS survey, we  imposed
similar criteria to those described by \cite {Steidel10}: (i) galaxy spectra
for the \textit{fg} and \textit{bg} galaxies must possess an accurately 
determined redshift ($95-100\%$; see \citealt{LeFevre15,Tasca17} for details), 
(ii) a redshift separation 0.1 $< z_{\rm{bg}}$ - $z_{\rm{fg}} \leqslant$ 1.0
between \textit{fg-bg} galaxy pairs to ensure that each spectrum contains a
significant common spectral coverage after shifting to the rest-frame of
the \textit {fg} galaxy and to avoid an overlap between the $fg$ CGM and the
$bg$ galaxy absorptions, and iii) a maximum projected angular separation of
23$\arcsec$.

\begin{figure}
\centering
\includegraphics[width=8.5cm]{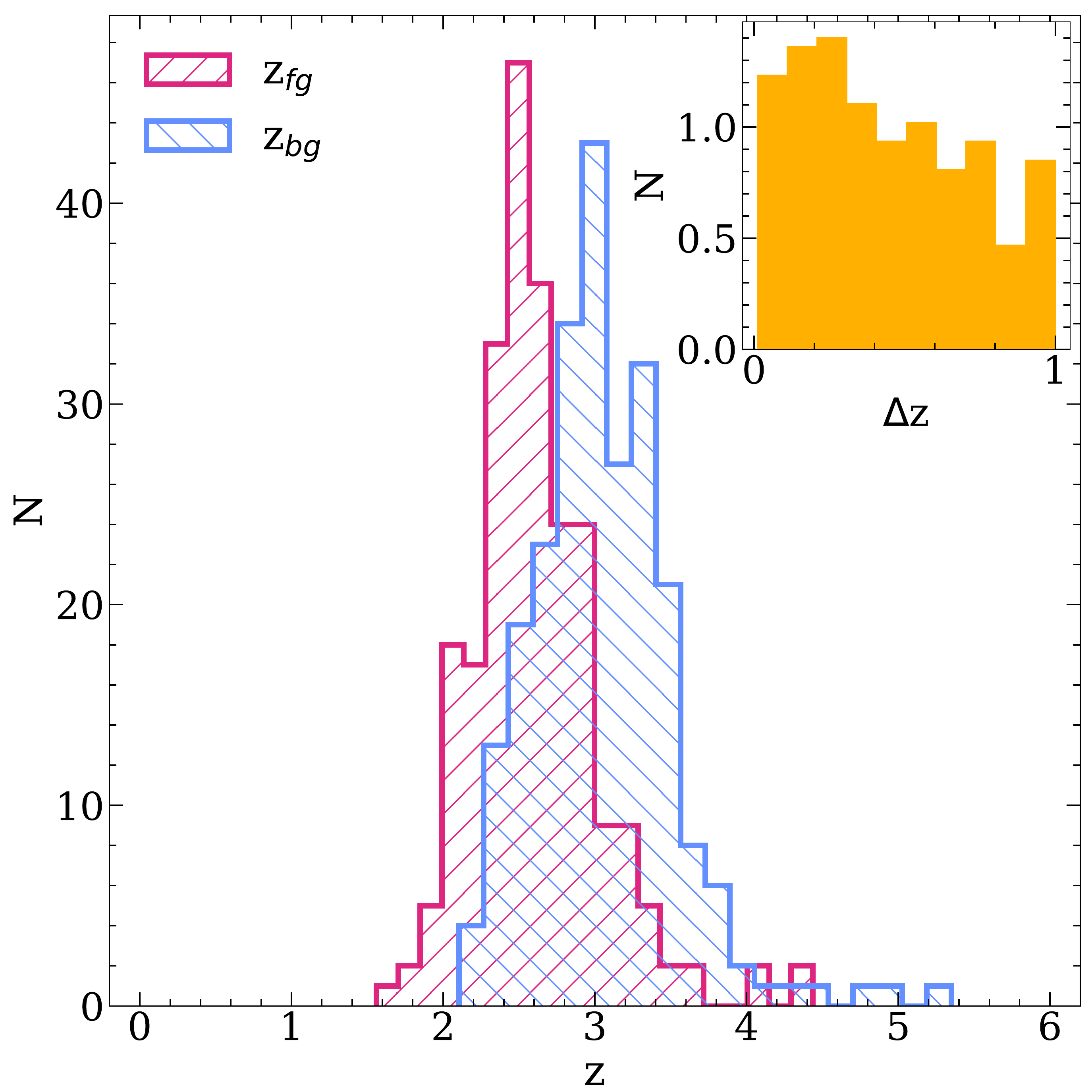}
\caption{Redshift distribution of the star-forming galaxy pairs 
(\textit{foreground} and \textit{background}) selected in this work.
The $\Delta$z = $z_{bg}$-$z_{fg}$ distribution is also included in the top right 
corner.}
\label{Fig:PairsHistZ}
\end{figure}  

Stacking the \textit{fg} spectra provides the  absorptions at the \textit{foreground}
galaxy's rest-frame. Figure \ref{Fig:PairsHistZ} shows the redshift
distributions of the \textit{foreground} and \textit{background} 
galaxy pairs and the \textit{fg-bg} pairs redshift
difference ($\Delta z=z_{bg} -z_{fg}$) distribution in the top right corner.
The spectroscopic sample includes 238 spectra within the redshift range 1.5
< z $\lesssim$ 4.43 ($\langle z \rangle$ = $2.6 \pm 0.41$) and is limited to
a 23$\arcsec$ maximum projected angular separation distance amongst galaxy
pairs that correspond to 187.2 kpc at z=2.6. This sample is then split into
four different bins according to the angular projected separation,
$\theta\leqslant11\farcs8$, $11\farcs8<\theta\leqslant16\farcs5$,
$16\farcs5<\theta\leqslant20\arcsec$, and $20\arcsec<\theta\leqslant
23\arcsec$, identified as samples S1, S2, S3, and S4, and were
defined in such a way that each bin contains approximately the same number
of galaxies (see Table \ref{tbl:PairsStat}) providing a comparable S/N for
each composite spectra. These angular projected separation bins correspond
to projected physical distances (hereafter impact parameters) $b$:
$\leqslant$ 68.6\,kpc, 68.6\,kpc $< b \leqslant$113.5\,kpc, 113.5\,kpc $<
b \leqslant$ 146.2\,kpc, and 146.2\,kpc $< b \leqslant$ 172.8\,kpc for
samples S1, S2, S3, and S4. We note that the conversion from
angular separation to physical impact parameter varies $\pm3\%$ over the
full redshift range $(1.5 < z < 4.5)$ of the \textit{foreground} galaxy
sample, thus in the following impact parameter $b$ and projected angular
separation $\theta$ will be used interchangeably. Table \ref
{tbl:PairsStat} contains a summary of the statistical properties of the
galaxy pair samples. \\ 

To generate our \textit{fg} and \textit{bg} composite spectra we co-added
individual spectra as follows. For each pair of galaxies, the spectrum of
the \textit{fg} was i) shifted to its own rest-frame; ii)
continuum-normalized using the full wavelength range\footnote{Full observed
wavelength coverage of $3650 < \uplambda < 9350 \, [\AA],$ which at
$z\sim$2.6 translates into a rest-frame wavelength coverage of $1013
< \uplambda < 2597 \, [\AA]$.}; iii) resampled\footnote{We   made use of {\sc
{pysynphot}} \citep{Blagorodnova14} to resample our spectra.} to a common
wavelength resolution $\Delta\uplambda$, defined by the maximum of the
shifted wavelength resolution ($\Delta\uplambda_{shf\textunderscore fg}$)
distribution of the galaxy sample used to generate the composite spectra
(e.g. $\Delta\uplambda\sim2\AA$ for $fg$ composite spectra as shown in
Figure \ref{Fig:Stk-Fg-Sep0-23}); and iv) smoothed with a Gaussian kernel
whose width size $\Delta\uplambda$ was defined before.

A similar approach was used to produce stacked spectra of the \textit
{bg} galaxies. Each individual \textit{bg} galaxy spectrum was i) shifted
into the \textit {fg} galaxy's rest-frame using the same systemic redshift
applied to their corresponding \textit{fg} galaxy spectrum; ii) continuum
normalized using the full wavelength range$^{1}$; iii)  resampled to a common
wavelength resolution $\Delta\uplambda$, defined by the maximum of the
shifted wavelength resolution($\Delta\uplambda_{shf\textunderscore bg}$)
distribution of the galaxy sample used to generate the composite spectra; iv)
smoothed with $\Delta\uplambda$ Gaussian kernel as defined before. The strong
interstellar absorption features  (see Fig 
\ref{Fig:Stk-Fg-Sep0-23} and Table \ref{Tbl:Lines}) located at the
redshift of each \textit{bg} rest-frame were masked because they  can
potentially contaminate the composite signal at the \textit
{fg} rest-frame. 

Finally, the \textit{foreground} and \textit{background} spectra were co-added
independently to produce composite stacked spectra. In all cases, for each
spectral bin, we calculated both the average and the median value, eventually
producing both an average and median co-added spectrum.


\section{Results}\label{sec:Res}

\begin{figure*}
   \centering
   \includegraphics[width=17cm]{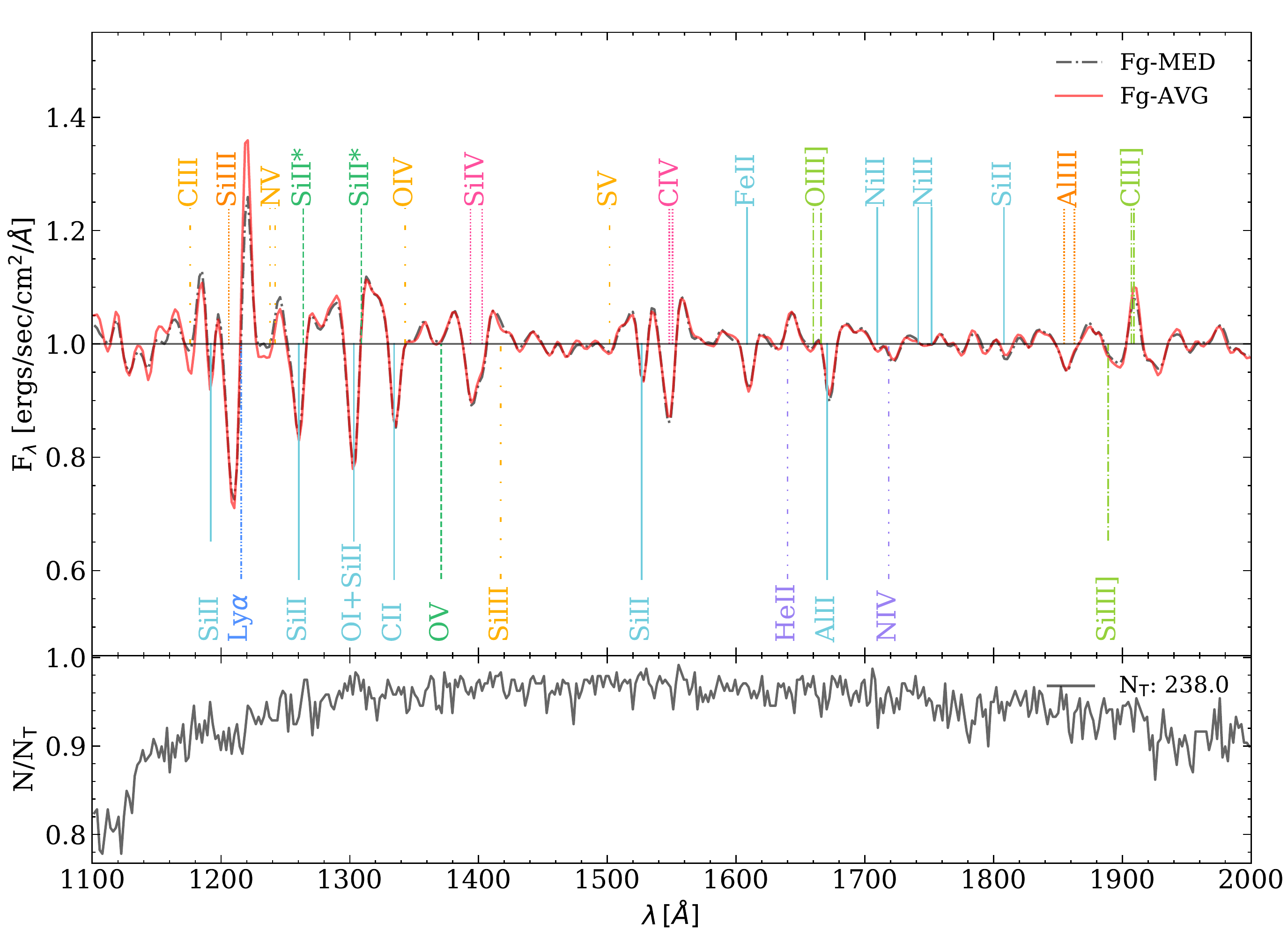}
   \caption{Composite UV spectra of 238 VUDS star-forming galaxies.
    Top: Median (black) and average (red) composite spectra shifted to
    rest-frame of 238 \textit{foreground} galaxies with projected angular
    separations $<$\,23$\arcsec$ ($b<187.2$\,kpc at z=2.6). Bottom: Fraction of contributing individual spectra to the composite spectra per
    unit wavelength. Several spectral lines of interest are indicated: H\,
    {\sc i} (Ly$\upalpha$) emission--absorption (blue),
    low-ionization state
    metal absorption lines (cyan), intermediate-ionization state metal
    absorption lines (peach), high-ionization state metal absorption lines
    (magenta), interstellar fine-structure emission line (green), absorption
    stellar photospheric lines (gold), emission nebular lines (lime), and
    emission--absorption lines associated with stellar winds (indigo).}
    \label{Fig:Stk-Fg-Sep0-23}
    \end{figure*}

Rest-frame UV spectra of star-forming galaxies at redshift $z\sim3$ are
commonly dominated by the emission of O and B stars; the CGM and/or IGM media
imprint absorption features on  this UV continuum
\citep{Sargent80,Bergeron86,Bergeron91,Lanzetta95}. Composite spectra can
provide different l.o.s. of the average absorption strength produced by the
gas located in these media
\citep{Adelberger05a,Shapley03}. Figure~\ref{Fig:Stk-Fg-Sep0-23} shows the
median and average composite spectra by stacking the 238 \textit
{foreground} galaxies. The lower panel shows a histogram of the fraction of
galaxies contributing to each spectral bin of the co-added spectrum. In our
analysis the separation zero ($b=0$) l.o.s. defines the interstellar medium
properties of the \textit{foreground} galaxies, while larger separations
($b>0$) l.o.s. define the properties of the CGM around the
\textit{foreground} galaxies at different separations.

While the l.o.s that we use to study the CGM cross through the whole CGM at a
given separation, the information about the ISM comes only from the
absorptions produced by the gas located in the half-galaxy that is the
closest to the observer. For this reason, to compare the strength of the
absorptions at separation 0 (the ISM) and at separation $>0$ (the
CGM), the former need to be corrected for this incompleteness.
Following \cite{Steidel10}, we apply the following factors: 1.45 for
low-ionization species (Si \,{\sc ii}, C \,{\sc ii}), 1.70 for Si \,{\sc iv},
and $\sim2$ for C \,{\sc iv}. The different l.o.s. probed by the \textit
{background} galaxy's light passing through the \textit{foreground} CGM help
us to trace neutral and ionized gas in H{\sc ii} star-forming regions and the
large-scale stellar winds produced by star formation activity \citep
{Shapley03}. The main spectral features identified in our \textit
{fg} composite spectra (Figure \ref{Fig:Stk-Fg-Sep0-23}) in the
1100-2000$\AA$ range are summarized in Table \ref{Tbl:Lines}.

\begin{table}
\centering
\caption{\label{Tbl:Lines}
Main spectral features observed in our VUDS stacked 
spectra in the 1100-2000 $\AA$ rest-frame range.}
\begin{tabular}{ccc}
  \\
  \hline\hline
  \multicolumn{1}{c}{Spectral line} &
  \multicolumn{1}{c}{$\lambdaup$ [$\AA$]} &
  \multicolumn{1}{c}{Type} \\
  \hline\hline  
C  \,{\sc iii} & 1176 & Photospheric \\
Si \,{\sc ii}  & 1192 & LIS absorption \\
Si \,{\sc iii} & 1206 & IIS absorption \\
Ly$\upalpha$   & 1215.7 & H \,{\sc i} \\
N \,{\sc v}    & 1238, 1242 & Photospheric \\ 
Si \,{\sc ii}  & 1260 & LIS absorption \\
Si \,{\sc ii*} & 1264.0 & Interstellar \\
O \,{\sc i}+Si \,{\sc ii}& 1303.2 & LIS absorption \\
Si \,{\sc ii*} & 1309.0 & Interstellar \\
C \,{\sc ii}   & 1334.5 & LIS absorption \\
O \,{\sc iv}   & 1343.0 & Photospheric \\
Si \,{\sc iv}  & 1393.8, 1402.8  & HIS absorption\\
Si \,{\sc iii} & 1417 & Photospheric \\
S \,{\sc v}    & 1501.8 & Photospheric \\
Si \,{\sc ii}  & 1526.7 & LIS absorption \\
C \,{\sc iv}   & 1548.2, 1550.8 & HIS absorption\\
Fe \,{\sc ii}  & 1608, & LIS absorption \\
Fe \,{\sc ii}  & 1610, & LIS absorption \\
He \,{\sc ii}  & 1640  & Stellar wind \\
O \,{\sc iii]} & 1660, 1666 & Nebular \\
Al \,{\sc ii}  & 1670.7  & LIS absorption \\
Ni \,{\sc ii}  & 1709.6 & LIS absorption \\
N \,{\sc iv}   & 1718.5 & Stellar wind \\
Ni \,{\sc ii}  & 1741.5, 1751.9 & LIS absorption \\
Si \,{\sc ii}  & 1808 & LIS absorption \\
Al \,{\sc iii} & 1854.7, 1862.7 & IIS absorption \\
Si \,{\sc iii]}& 1889 & Nebular \\
C \,{\sc iii]} & 1908.7 & Nebular \\
\hline
\end{tabular}
\tablebib{
\citet{LeFevre15,Halliday08,Shapley03}; OTELO (OSIRIS Tunable Emission Line Object survey \url{http://research.iac.es/proyecto/otelo/pages/data-tools/spectral-line-summary.php}).}
\end{table}

\begin{table*}
\centering
\caption{\label{tbl:PairsStat}
\textit{Foreground--background} (\textit{fg-bg}) galaxy pair statistics 
according to the projected angular separation ($\theta$).}
\begin{tabular}{cccccccccc}
  \\
  \hline\hline
  \multicolumn{1}{c}{ID} &
  \multicolumn{1}{c}{N} &
  \multicolumn{1}{c}{$\theta$} & 
  \multicolumn{1}{c}{$\langle \theta \rangle$} &
  \multicolumn{1}{c}{$b$ (kpc)} & 
  \multicolumn{1}{c}{$\langle b \rangle$ (kpc)} & 
  \multicolumn{1}{c}{$z_{\rm{fg}}$} &
  \multicolumn{1}{c}{$\langle \rm{z}_{\rm{fg}}\rangle$} &
  \multicolumn{1}{c}{$z_{\rm{bg}}$}  & 
  \multicolumn{1}{c}{$\langle \rm{z}_{\rm{bg}}\rangle$} \\
  \multicolumn{1}{c}{c1} & 
  \multicolumn{1}{c}{c2} &
  \multicolumn{1}{c}{c3} &
  \multicolumn{1}{c}{c4} &
  \multicolumn{1}{c}{c5} &
  \multicolumn{1}{c}{c6} &
  \multicolumn{1}{c}{c7} &
  \multicolumn{1}{c}{c8} &
  \multicolumn{1}{c}{c9} &
  \multicolumn{1}{c}{c10} \\
  \hline\hline  
ALL & 238 &  <23$\arcsec$             & 15$\farcs$7 $\pm$ 4$\farcs$9 &  13.8 - 187.2 & 125.0 $\pm$ 39.5 & 1.5-4.4 & 2.60 & 2.1-5.3 & 3.04 \\
S1  &  59 &  <11$\farcs$8             &  8$\farcs$7 $\pm$ 2$\farcs$0 &  13.8 -  97.8 & 68.6  $\pm$ 16.3 & 1.5-4.3 & 2.55 & 2.1-4.9 & 2.98 \\
S2  &  62 &   11$\farcs$8-16$\farcs$5 & 14$\farcs$4 $\pm$ 1$\farcs$4 &  92.6 - 135.9 & 113.5 $\pm$ 12.8 & 1.7-4.3 & 2.62 & 2.1-4.9 & 3.08 \\
S3  &  60 &   16$\farcs$5-20$\arcsec$ & 18$\farcs$4 $\pm$ 0$\farcs$9 & 127.1 - 164.6 & 146.2 $\pm$  8.8 & 1.9-3.6 & 2.66 & 2.3-4.5 & 3.11 \\
S4  &  57 &   20$\arcsec$-23$\arcsec$ & 21$\farcs$5 $\pm$ 0$\farcs$8 & 146.7 - 187.2 & 172.8 $\pm$  8.4 & 1.8-4.4 & 2.59 & 2.1-5.3 & 3.02 \\
  \hline\hline
\end{tabular}
\tablefoot{
Column 1: Sample ID; 
Column 2: number of galaxies per sample; 
Column 3: range of angular separation of the \textit{fg-bg} pairs;
Column 4: mean angular separation of the \textit{fg-bg} pairs within the
$\theta$ range;
Column 5: range of the impact parameter in kpc assuming $z_{med}$ within the
projected separation bin for conversion;
Column 6: mean impact parameter within the $\theta$ range in kpc;
Columns 7 and 8: redshift ranges and mean redshift of the \textit
{foreground} galaxies sample within their corresponding $\theta$ range, and
Columns 9 and 10: redshift ranges and mean redshift of the \textit
{background} galaxies sample within their corresponding $\theta$ range.}
\end{table*}

In the 900--1900\,$\AA$ range we   explored the following absorption lines:
Ly$\upalpha$ ($\uplambda$ 1215.7 $\AA$), 
O\,{\sc i}+Si\,{\sc ii} ($\uplambda\uplambda$ 1303.2 $\AA$), 
C\,{\sc ii} ($\uplambda$ 1334.5 $\AA$), 
Si\,{\sc iv} ($\uplambda\uplambda$ 1393, 1402 $\AA$), 
Si\,{\sc ii} ($\uplambda$ 1526.7 $\AA$), 
C\,{\sc iv} ($\uplambda\uplambda $ 1548.2, 1550.8 $\AA$), 
Fe\,{\sc ii} ($\uplambda$ 1608.5 $\AA$), 
Al\,{\sc ii} ($\uplambda$ 1670.8 $\AA$), and 
Al\,{\sc iii} ($\uplambda$ 1862.8 $\AA$).
First, we present their equivalent widths as a function of the angular
separation between galaxy pairs ($\theta$) (see Sect. \ref{sec:EW-Rad}). We
then examine the correlations between the strength of the  Ly$\upalpha$, LIS (C\,
{\sc ii}, Si\,{\sc ii}), and HIS (C\,{\sc iv}, Si\,{\sc iv}) lines as a
function of different galaxy properties, including stellar mass, star
formation rate (see Sect.~\ref{sec:EW-SFR}), effective radius ($r_{\rm
{eff}}$), and the azimuthal angle ($\phi$) (see Sect.~\ref{sec:EW-Mrph}). We
also explore the C\,{\sc ii}/C\,{\sc iv} line ratio (see Sect.~\ref
{sec:LR}) as a function of the projected angular separation, stellar mass,
and star formation rate.

\subsection{Radial dependence}\label{sec:EW-Rad}

\begin{figure*}
\centering
\includegraphics[width=17cm]{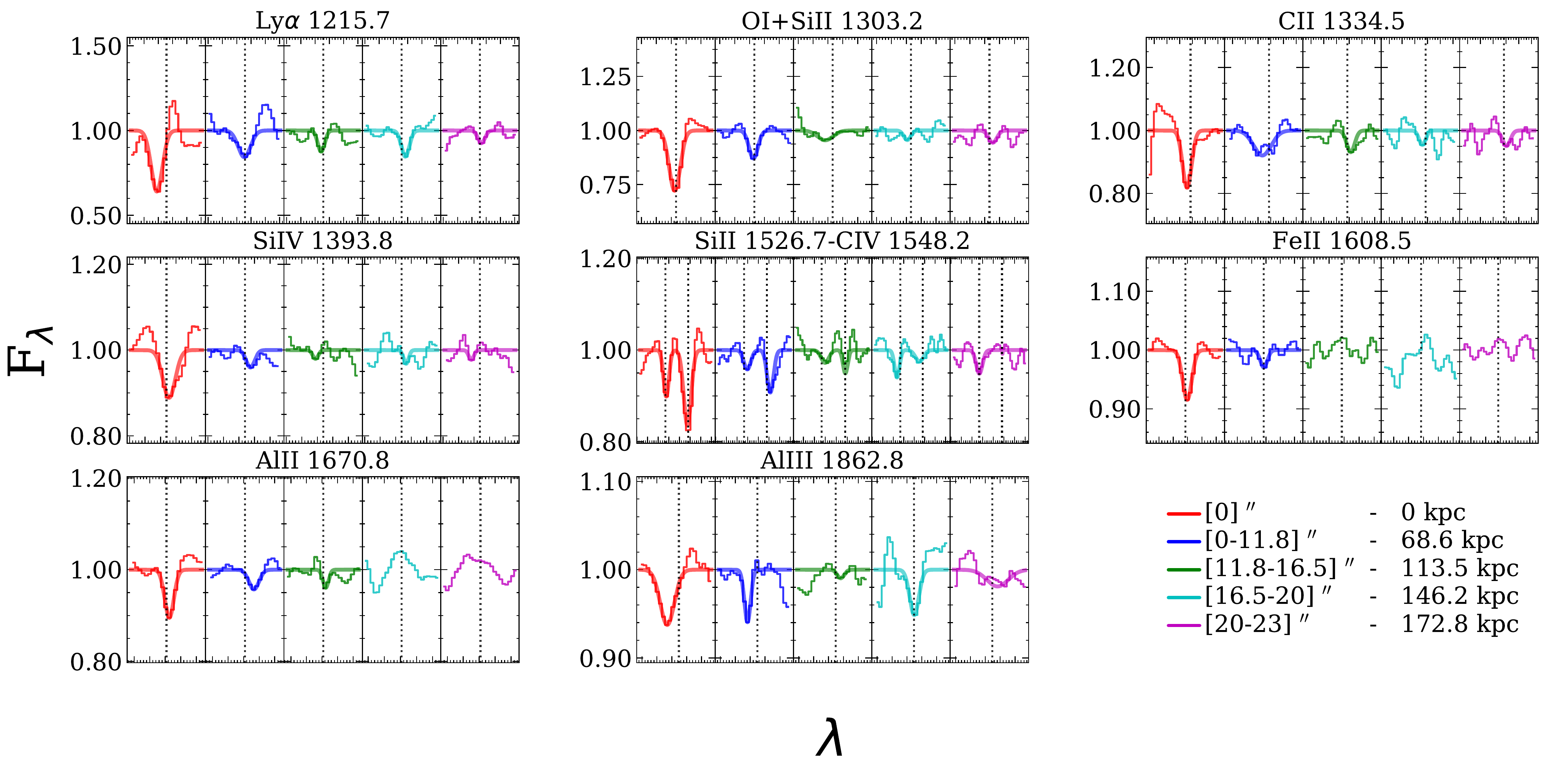}
  \caption{Median absorption lines (thin lines) detected from 
    our 238 \textit{foreground--background} galaxy pairs, including
    all \textit{foreground} galaxies (red) and \textit{background} galaxies
    split by their projected angular separations:
    <11$\farcs$8 (blue), 
    11$\farcs$8-16$\farcs$5 (green), 
    16$\farcs$5-20$\arcsec$ (cyan), and 
    20$\arcsec$-23$\arcsec$ (magenta). The  
    bold lines indicate
    the Gaussian profile fit to measure the 
    strength of the line absorptions (see Sect. \ref{sec:EW-Rad}).}
\label{Fig:Lns-Bg-Sep-4}
\end{figure*}

Here we present absorption lines produced by gas located in the CGM of
star-forming galaxies at $z\sim2.6$. Figure~\ref{Fig:Lns-Bg-Sep-4} shows the
average absorption spectra obtained after stacking the \textit
{background} galaxy spectra at the redshift of the
\textit{foreground} galaxy as a function of their projected angular
separation: $<11\farcs8$, $11\farcs8$--$16\farcs5$,
$16\farcs5$--$20\arcsec$, and $20\arcsec$--$23\arcsec$. We note that as a
consequence of the  VUDS design and selection criteria, the number of
galaxies with small projected angular separations ($<6\arcsec$) is scarce.
The strengths of the line absorptions were measured by fitting a single
Gaussian profile to the stacked spectra. Equivalent widths ($W_{0}$) of each
line profile are obtained by integrating the Gaussian fit of each stacked
spectrum around the central wavelength of the Gaussian fit to the line using
an integration window within the $\pm5\sigma$ range.

All equivalent width measurements are given in the rest-frame and we use
positive (negative) equivalent widths to indicate absorption (emission). The
errors on the equivalent width measurements are determined using a bootstrap
approach. For each set of galaxy \textit{background} spectra, a thousand
co-added spectra are generated from random selections, with replacements from
that same sample in order to   preserve the number of evaluated sources. For
each of these co-adds, the line absorption equivalent widths are estimated.
The 16th and 84th percentiles of the distribution of equivalent widths are
taken as error intervals for the original measurement. We note that owing to
the nature of star-forming galaxies, which show a diversity of spectral
features(e.g. ISM absorption lines, photospheric stellar and nebular lines,
and Ly$\upalpha$ emission), the absorption lines in our composite spectra
show both an emission and absorption component. The emission components can
be produced by the same transition responsible for producing the absorption
line (e.g. Ly$\upalpha$), produced by nearby fine-structure transitions, or
in some cases they can be completely absent.

\begin{figure*}
   \centering
   \includegraphics[height=21cm]{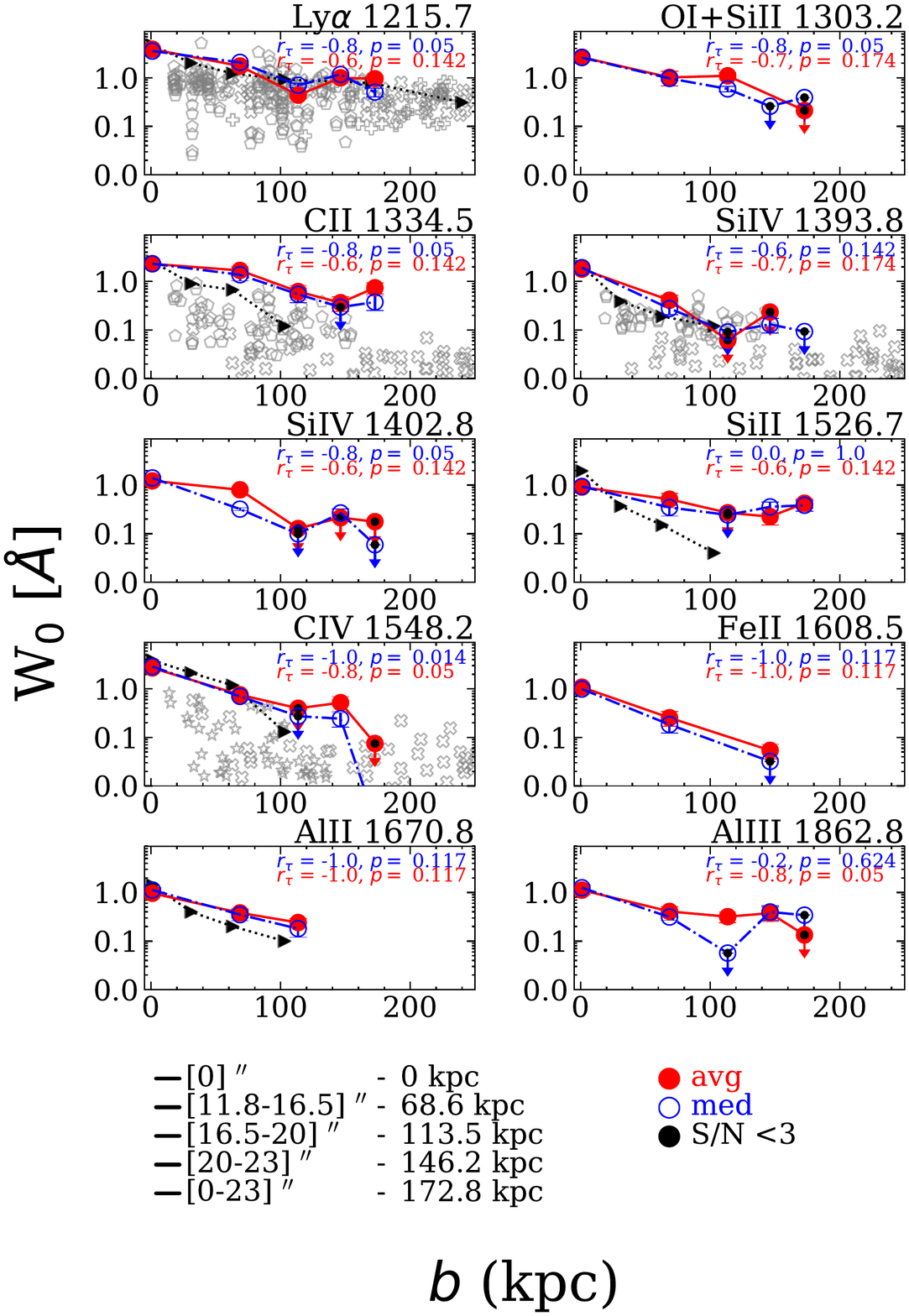}
   \caption{
   Rest equivalent width ($W_{0}$) as a function of the impact 
   parameter ($b$) obtained from the line profiles in Fig. \ref{Fig:Lns-Bg-Sep-4},
   corresponding to the \textit{foreground} composite spectra (0 kpc) and the
   \textit{background} composite spectra at $\langle b \rangle$
   68.6 kpc (8$\farcs$7), 
   113.5 kpc (14$\farcs$4), 
   146.2 kpc (18$\farcs$4), and 
   172.8 kpc (21$\farcs$5). 
   Average and median $W_{0}$ are respectively shown in filled red and open blue symbols;
   solid black symbols correspond to  upper limits with S/N
   <3. Reported values from the literature including low-redshift values
   (open grey symbols):
   \citealt{Bordoloi18} (stars), 
   \citealt{Borthakur15} (pentagons), 
   \citealt{Liang14} (crosses), 
   \citealt{Prochaska11}  (diamonds), 
   \citealt{Werk13} (pluses), and  
   high-redshift values \citealt{Steidel10} (black triangles). 
   The error bars correspond to 1$\sigma$ confidence intervals for average
   (red filled circles) or median (blue open circles) values based on a
   bootstrap analysis. Also included are  the results from the Kendall-Tau
   correlation test: the correlation coefficients $r_{\tau}$ and the   corresponding $p$-value (probability of no correlation).}
   \label{Fig:EW-Pfl-Bg-sep-4}
   \end{figure*}

Figure \ref{Fig:EW-Pfl-Bg-sep-4} shows the radial profiles of the absorption
rest-frame equivalent width obtained from the average (red) and median
(blue) stacked absorption specrta presented in Figure~\ref
{Fig:Lns-Bg-Sep-4}. Median composite spectra are expected to be free of
artefacts and contamination such as sky residuals, unexpected absorption or
emission features in the \textit{background} galaxy spectra. On the other
hand, the average composite spectra could be affected by strong absorption
lines coming from individual l.o.s. Median and average composite spectra are
in general concordant, and trace similar trends. We also tested other
combination techniques, including straight average, average sigma-clipping,
and average continuum-weighting, which produce very similar composite
spectra. Table \ref{tbl:EW-SEP} contains a summary of the measured equivalent
widths as a function of the projected angular separation, including the
significance (in terms of S/N) of the detections. Noise estimations were
computed with the 
{\sc DER$\_$SNR}\footnote{\url{http://www.stecf.org/software/ASTROsoft/DER_SNR/}} 
algorithm  \citep{Stoehr08} on a continuum band-pass adjacent to each central
absorption line. We limit our analyses to detections with S/N\,$\geq$\,3;
however, upper limits with $S/N\,<\,3$ are also included (highlighted using
solid black symbols and arrows in Figure \ref{Fig:EW-Pfl-Bg-sep-4}) as these
detections are useful to outline the possible correlations and/or trends of
the profiles. We note that some precautions should be taken when interpreting
the absorption strengths of these spectral features as they might not reflect
the true conditions of the cold and hot gas CGM components. For instance, the
balances amongst singly ionized species of some elements (e.g. Al, Cl) can be
altered by dielectronic recombination \citep{Black73,Watson73,Jura74} or
ion-molecule reactions (charge exchange reactions of ionized species with
neutral hydrogen and helium \citealt{Steigman75b}). While the abundances
relative to neutral hydrogen of some elements (e.g. O, N, P, Cl, Si, Cr, Mn,
Fe, Ni) can be affected by factors of 2-10, Al is depleted  by a factor of
$\sim$ 100 \citep{York79}, making Al\,{\sc ii} a biased tracer of the neutral
gas phase. In addition, Fe\,{\sc ii} is contaminated by Fe\,{\sc iv}
($\uplambda$ 1610 $\AA$) and Fe\,{\sc ii} ($\uplambda$ 1611.5 $\AA$)
lines \citep{Judge92,Shull83,Pickering02}, making it difficult to determine
accurately its absorption strength. Subsequently, doubly ionized species
(e.g. Si\,{\sc iii}, Al\,{\sc iii}) are intermediate-ionization state
(IIS) tracers of moderately photoionized warm gas sensitive to both diffuse
ionized gas (traced by HIS) and to  denser partly neutral gas. Depending on
the species, they can be correlated with cold gas components (e.g. Si\,
{\sc iii} \citealt{Richter16}) or hot gas components  (e.g. Al\,
{\sc iii} \citealt{Vladilo01}), and they can be used to infer the relative
mix of neutral and ionized material \citep{Howk99b}.

We note that the composite spectra considering \textit{foreground}
(down-the-barrel) galaxies (from our $fg-bg$ galaxy pairs split by
separation) do not show any significant dependency on the $W_{0}$ of
absorption lines with impact parameter $b$;  these absorption lines show an
average variance of $\leq15\%$ amongst the absorption lines included in \ref
{Fig:EW-Pfl-Bg-sep-4}. This is in agreement with the $< 10\%$ variance
that \cite{Steidel10} reported for their brightest absorption lines.
Considering the \textit{background} galaxies from our $fg-bg$ galaxy pairs
split by separation, our radial profiles show a general negative gradient, as
reported by previous works. As the impact parameter increases, the strength
of the detected absorption feature decreases. This is the case for all the
spectral lines considered in our analysis. Recently, \cite
{Kacprzak21} compiled measurements of low-ionization state absorption lines
(O\,{\sc i}, C\,{\sc ii}, N\,{\sc ii}, Si\,{\sc ii}),  and high-ionization
state absorption lines(C\,{\sc iii}, N\,{\sc iii}, Si\,{\sc iii}, C\,
{\sc iv}, Si\,{\sc iv}, N{\sc v}, and O{\sc vi}) in the CGM of low-redshift
($z \lesssim 0.3$) galaxies with $9\leq \rm{log[M}_{\bigstar}/\rm{M}_
{\odot}]\leq11$, using quasars as
\textit{background} spotlights. Measurements of low-redshift galaxies
\citep{Bordoloi18,Borthakur15,Johnson17,Liang14,Prochaska11,Werk13} and
high-redshift galaxies ($z \sim 2.3$)  \citep{Steidel10} are also
overplotted in Figure~\ref{Fig:EW-Pfl-Bg-sep-4}. 

Although at low redshift these lines can be detected at distances of up to
300\,kpc, at high redshift using galaxy--galaxy pairs, 
\cite{Steidel10} reported detections that are limited to 100 kpc; they
included 42 additional bright quasars--galaxy pairs that provide access to
the low-density gas in the CGM, and  extended their Ly$\upalpha$ detections
up to 280\,kpc. More recently, using galaxy images at $\langle
z \rangle \sim$\,2.4,
\cite{Chen20} measured the Ly$\upalpha$ excess relative to the \textit
{background} intergalactic medium, probing the CGM gas up to impact
parameters of 2000\,kpc. Here, for star-forming galaxies at a mean redshift
$\langle z \rangle$ = 2.6, we report detections at distances up to 146\,kpc
and 172\,kpc  for HIS and LIS--Ly$\upalpha$. These results include Al\,
{\sc iii}, which is detected for the first time in the CGM of high-redshift
star-forming galaxies up to distances of $\sim$150\,kpc. Low-ionization
state lines (C\,{\sc ii}, Si\,{\sc ii}, Al\,{\sc ii}) show less steep radial
profiles compared with high-ionization lines. The LIS radial profile shows
an abrupt decay at smaller radii. From separation zero to $\langle
b \rangle\sim 68$\,kpc, the strengths of C\,{\sc ii}, Si\,{\sc ii}, and Al\,
{\sc ii} are reduced by a factor of 1.7, 2.5, and 3.2, while Si\,{\sc iv}
and C\,{\sc iv} are reduced by a factor of 6.8 and 4.5.

\cite{Steidel10} reported that at a separation of $\langle b \rangle$ =
103\,kpc, lines can be hardly detected in their stacked spectra,
low-ionization state metal absorption lines cannot be detected, and C\,
{\sc iv} is detected with marginal significance. Here we are able to detect
LIS (C\,{\sc ii} and Si\,{\sc ii}) and HIS (Si\,{\sc iv}, C\,{\sc iv}) both
with S/N$\geq3$ up to $\langle b \rangle=$\,172\,kpc and 146\,kpc
separations. We note that beyond this separation
($\theta>23\arcsec$=172\,kpc) we are unable to detect any significant signal
from the \textit{background} composite spectra. This might be caused by
differences in the S/N and resolution of the individual spectra used in the
two analyses. The spectral line features considered here are more difficult
to detect in our low-resolution spectra compared to  high-resolution spectra
at the same S/N level. Higher-resolution spectra could  resolve spectral
lines that are blended(e.g. O\,{\sc i}--Si\,{\sc ii}: $\uplambda\uplambda$
1303, 1307 $\AA$, Si\,{\sc iv}: $\uplambda\uplambda$ 1393, 1402 $\AA$) or
that are affected by close photospheric, interstellar, or nebular spectral
lines(e.g. Fe\,{\sc ii}, Fe\,{\sc iv}: $\uplambda\uplambda 1608, 1610 \AA$,
Si\,{\sc ii}, Si\,{\sc ii}*: $\uplambda\uplambda 1260, 1264 \AA$),
disentangling the spectral features and thus providing  more accurate $W_
{0}$ measurements. Another possibility to explain these discrepancies is that
they might be the result of differences in the  S/N of both parent dataset
samples. We know that the two parent samples adopted opposite observational
strategies. On the one hand, 
\cite{Steidel04} deliberately kept  the total exposure times short on their
observations to maximize the number of galaxies for which redshifts could be
measured. This led to a dataset with spectral quality (S/N) that varies
considerably amongst their spectra \citep{Steidel04,Steidel10}, and that
could be affecting their redshift determinations. On the other hand, VUDS
provides a homogeneous dataset with integration times of $\simeq $14h per
target, reaching  a S/N on the continuum at 8500 $\AA$ of S/N=5 \citep
{LeFevre15}. Moreover, in this work we   consider galaxies with reliability
flags ($z_{f}$ = 3, 4) with 95-100\% probability of being correct. However,
it is not clear if similar constraints on their redshift determinations were
adopted by \cite{Steidel10} to define their galaxy--galaxy pair sample. This
is crucial, as composite spectra can be affected by spectral offsets in the
individual spectra. LIS and HIS spectral absorptions could be washed out or
artificially boosted by considering individual spectra with low S/N and
low-reliability redshift determinations to generate composite spectra. The
spectra considered to generate our stacked spectra may show an average S/N
higher than the average S/N of the spectra considered by \cite
{Steidel10}. If similar separations $\langle b\rangle $ are considered, for
example  Steidel's P3 sample at $\langle b\rangle =103$ kpc
($\langle \theta \rangle$ = 12$\farcs$5) and this work's S2 sample $b=113.5$
kpc ($\langle \theta \rangle$ = 14$\farcs$4), Steidel's P3 sample considers
approximately five times the number of objects (N=306) considered in our S2
sample (N=62). This would imply that our spectra have an average individual
S/N that is higher than  the average S/N of the dataset used by \cite
{Steidel10}, allowing us to detect similar or weaker spectral absorptions at
larger separations.

To assess the correlation between $W_{0}$ and $b$, we implemented a
Kendall-Tau correlation test. The correlation coefficients $r_{\tau}$ and the
corresponding $p$-value (the probability of no correlation) are provided in
the same figure.  In all cases we find a robust anticorrelation of the
strength of the absorption line as a function of the impact parameter; in
particular, Ly$\upalpha$, C\,{\sc ii}, C\,{\sc iv}, and Fe\,{\sc ii} all show
a strong correlation ($p<0.05$), while we find that O\,{\sc i}, Si\,{\sc iv},
Si\,{\sc ii}, Al\,{\sc ii}, and Al\,{\sc iii} present a flatter radial
profile, which  results in a lower significance of the anticorrelation
($p\sim0.1$). To assess the scale of the relationship between $W_{0}$ and
$b$, we computed the slopes of the radial profiles shown in Fig \ref
{Fig:EW-Pfl-Bg-sep-4}. We find that Ly$\alphaup$ shows a slope of
-1.20. $\pm$ 0.01; LIS (O\,{\sc i}+Si\,{\sc ii}, C\,{\sc ii}, Si\,{\sc ii},
Fe\,{\sc ii}, and Al\,{\sc ii)} absorption lines show slopes of  -1.34 $\pm$
0.01, -1.31  $\pm$  0.01, -1.53  $\pm$  0.02, -1.50  $\pm$  0.07, and -1.44
$\pm$ 0.05, respectively;  Al\,{\sc iii} shows a slope of  -1.38 $\pm$ 0.06;
and HIS (Si\,{\sc iv} and C\,{\sc iv)}  absorption lines show slopes
of -1.10  $\pm$ 0.04 and -1.24 $\pm$  0.01. If we focus on low- and
high-ionization lines of the same species, and consider Si\,{\sc ii}--Si\,
{\sc iv} and C\,{\sc ii}--C\,{\sc iv} absorption line pairs, we find that
the slopes of these absorption lines are different at $\geq5\sigma$ level.
Compared with the  \cite{Steidel10} radial profiles, our results show slopes
that are different at a 5$\sigma$ level for Si\,{\sc ii} and at a  3$\sigma$
level for C\,{\sc ii}  and Al\,{\sc ii} absorption lines. These differences
suggest that within the CGM, cold and dense gas is more spatially extended
in galaxies at $z\sim2.6$ compared with galaxies at $z\sim2.3$ and lower
redshifts, as probed by Si\,{\sc ii} C\,{\sc ii}, in agreement with the
expected higher covering factors of neutral gas at higher redshifts \citep
{Reddy16,Du18,Sanders21}. Compared with low-redshift CGM studies
(see Figure~\ref{Fig:EW-Pfl-Bg-sep-4}), our Ly$\upalpha$, LIS, and HIS
rest-frame equivalent width radial profiles are at the upper envelope of the
equivalent width measurements at lower redshifts, suggesting a potential
redshift evolution for the CGM gas content that produces these absorptions.

\begin{table*}
\centering
\caption{\label{tbl:EW-SEP} Median absorption line strengths measured ($W_
 {0} \, \AA$) in stacked spectra as a function of the average impact
 parameter ($\langle b \rangle$).}
\begin{tabular}{ccccccccccccccccccccccccccccccccccccccccc}
  \\
  \hline\hline
  \multicolumn{1}{c}{$\langle b \rangle$ (kpc)}  &
  \multicolumn{1}{c}{$W_{0}$ [$\AA$]}&
  \multicolumn{1}{c}{S/N} &
  \multicolumn{1}{c}{$W_{0}$ [$\AA$]}&
  \multicolumn{1}{c}{S/N} &
  \multicolumn{1}{c}{$W_{0}$ [$\AA$]}&
  \multicolumn{1}{c}{S/N} &
  \multicolumn{1}{c}{$W_{0}$ [$\AA$]}&
  \multicolumn{1}{c}{S/N} &
  \multicolumn{1}{c}{$W_{0}$ [$\AA$]}&
  \multicolumn{1}{c}{S/N} \\
  \multicolumn{1}{c}{c1} &
  \multicolumn{1}{c}{c2} &
  \multicolumn{1}{c}{c3} &
  \multicolumn{1}{c}{c4} &
  \multicolumn{1}{c}{c5} &
  \multicolumn{1}{c}{c6} &
  \multicolumn{1}{c}{c7} &
  \multicolumn{1}{c}{c8} &
  \multicolumn{1}{c}{c9} &
  \multicolumn{1}{c}{c10} &
  \multicolumn{1}{c}{c11} \\
  \hline\hline
  &
  \multicolumn{2}{c}{Ly$\upalpha$ 1215.7} &
  \multicolumn{2}{c}{O\,{\sc i}--Si\,{\sc iii} 1303.2} &
  \multicolumn{2}{c}{C\,{\sc ii} 1334.5} &
  \multicolumn{2}{c}{Si\,{\sc iv} 1393.8} &
  \multicolumn{2}{c}{Si\,{\sc iv} 1402.8} \\
  \hline\hline  
  0.0   & 3.54 $\pm$ 0.02 & 58 & 2.61 $\pm$ 0.15 & 17  & 2.31 $\pm$ 0.08 & 20 & 1.91 $\pm$ 0.01 & 18  & 1.40 $\pm$ 0.05 & 67\\
  68.6  & 1.71 $\pm$ 0.13 & 14 & 0.97 $\pm$ 0.10 &  9  & 1.37 $\pm$ 0.02 &  4 & 0.28 $\pm$ 0.06 &  4  & 0.32 $\pm$ 0.06 & 15\\
  113.5 & 0.75 $\pm$ 0.05 &  9 & 0.59 $\pm$ 0.07 &  8  & 0.54 $\pm$ 0.01 &  3 & 0.09 $\pm$ 0.04 &  2  &      <0.10     & <2\\
  146.2 & 1.08 $\pm$ 0.13 & 13 &      <0.26      & <2  & 0.30 $\pm$ 0.09 &  2 & 0.13 $\pm$ 0.19 &  3  & 0.27 $\pm$ 0.31 & 5 \\
  172.8 & 0.46 $\pm$ 0.14 &  5 &      <0.39      & <2  & 0.38 $\pm$ 0.10 &  3 &      <0.09      & <2  &      <0.06     & <2\\
  \hline\hline
  &
  \multicolumn{2}{c}{Si\,{\sc ii}  1526.7} &
  \multicolumn{2}{c}{C\,{\sc iv}   1548.2} &
  \multicolumn{2}{c}{Fe\,{\sc ii}  1608.5} &
  \multicolumn{2}{c}{Al\,{\sc ii}  1670.8} &
  \multicolumn{2}{c}{Al\,{\sc iii} 1862.8} \\  
  \hline\hline
  0.0   & 0.94 $\pm$ 0.02 & 18 & 2.87 $\pm$ 0.11 & 20  & 0.99 $\pm$ 0.01 &  6  & 1.13 $\pm$ 0.03 & 10  & 1.25 $\pm$ 0.01 &  8   \\
  68.6  & 0.35 $\pm$ 0.03 & 3  & 0.70 $\pm$ 0.04 & 13  & 0.18 $\pm$ 0.03 &  3  & 0.35 $\pm$ 0.03 &  4  & 0.31 $\pm$ 0.11 &  7   \\
  113.5 & 0.24 $\pm$ 0.09 & 2  & 0.27 $\pm$ 0.04 & 2   &       <0.03     &  <2 & 0.18 $\pm$ 0.03 &  3  &       <0.06     & <2  \\
  146.2 & 0.36 $\pm$ 0.05 & 4  & 0.24 $\pm$ 0.02 & 3   &       ----      & --- &       ----      & --- & 0.40 $\pm$ 0.02 &  3   \\
  172.8 & 0.39 $\pm$ 0.04 & 4  &      <0.05      & <2  &       ----      & --- &       ----      & --- &       <0.34     & <2  \\
\hline\hline 
\end{tabular}
\tablefoot{
\textit{Foreground--background (fg-bg)} galaxy pairs split by their average impact parameters 
($\langle b \rangle$):
68.6\,kpc (N=60), 
113.5\,kpc (N=62), 
146.2\,kpc (N=60), and
172.8\,kpc (N=57). 
The errors on the equivalent width ($W_{0}$) measurements 
correspond to 1$\sigma$ confidence intervals 
based on a bootstrap analysis (see Sect. \ref{sec:EW-Rad}).
}
\end{table*}


\subsection{Star formation and stellar mass dependence}\label{sec:EW-SFR}

\begin{figure*}
\begin{tabular}{cc}
  \includegraphics[width=85mm]{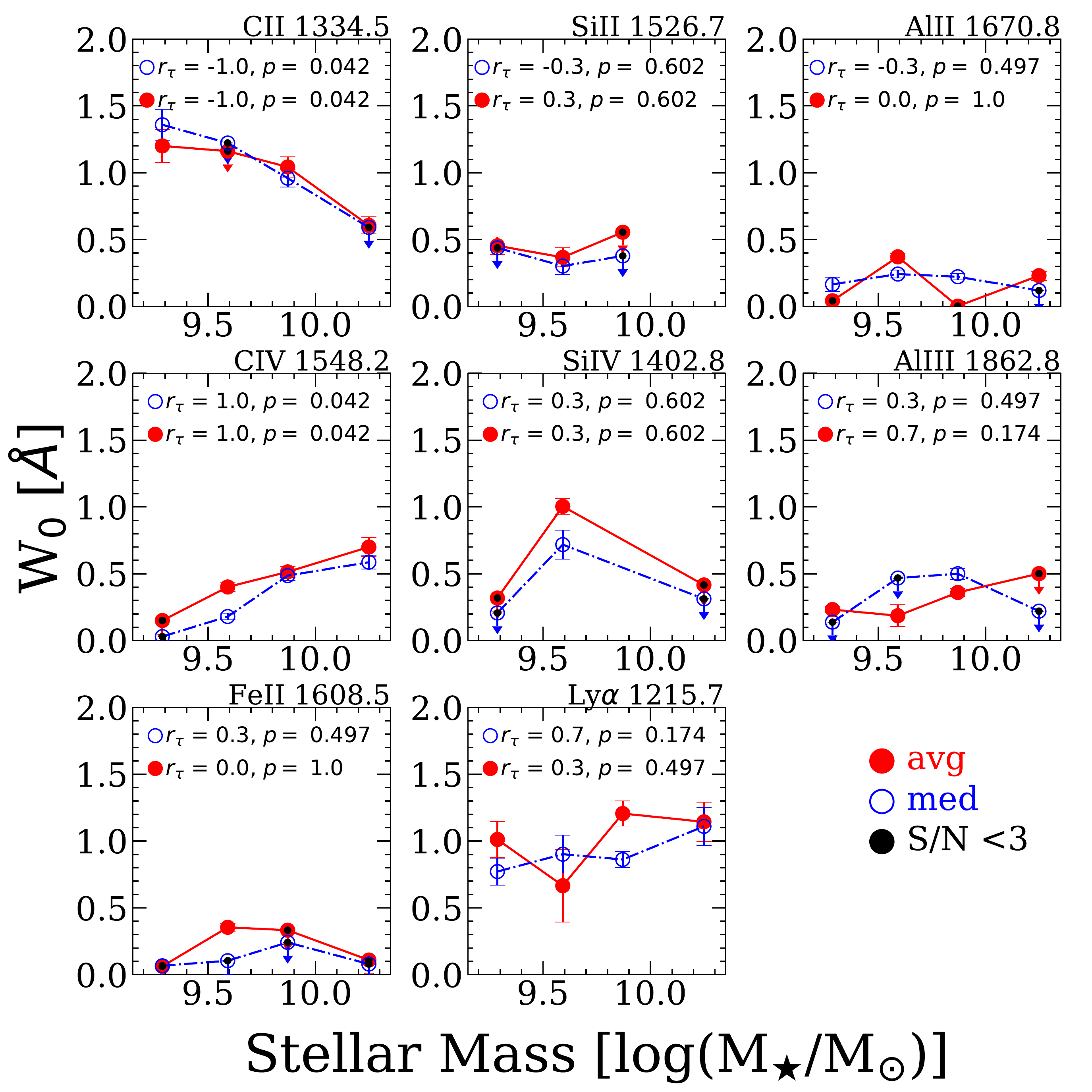} &
  \includegraphics[width=85mm]{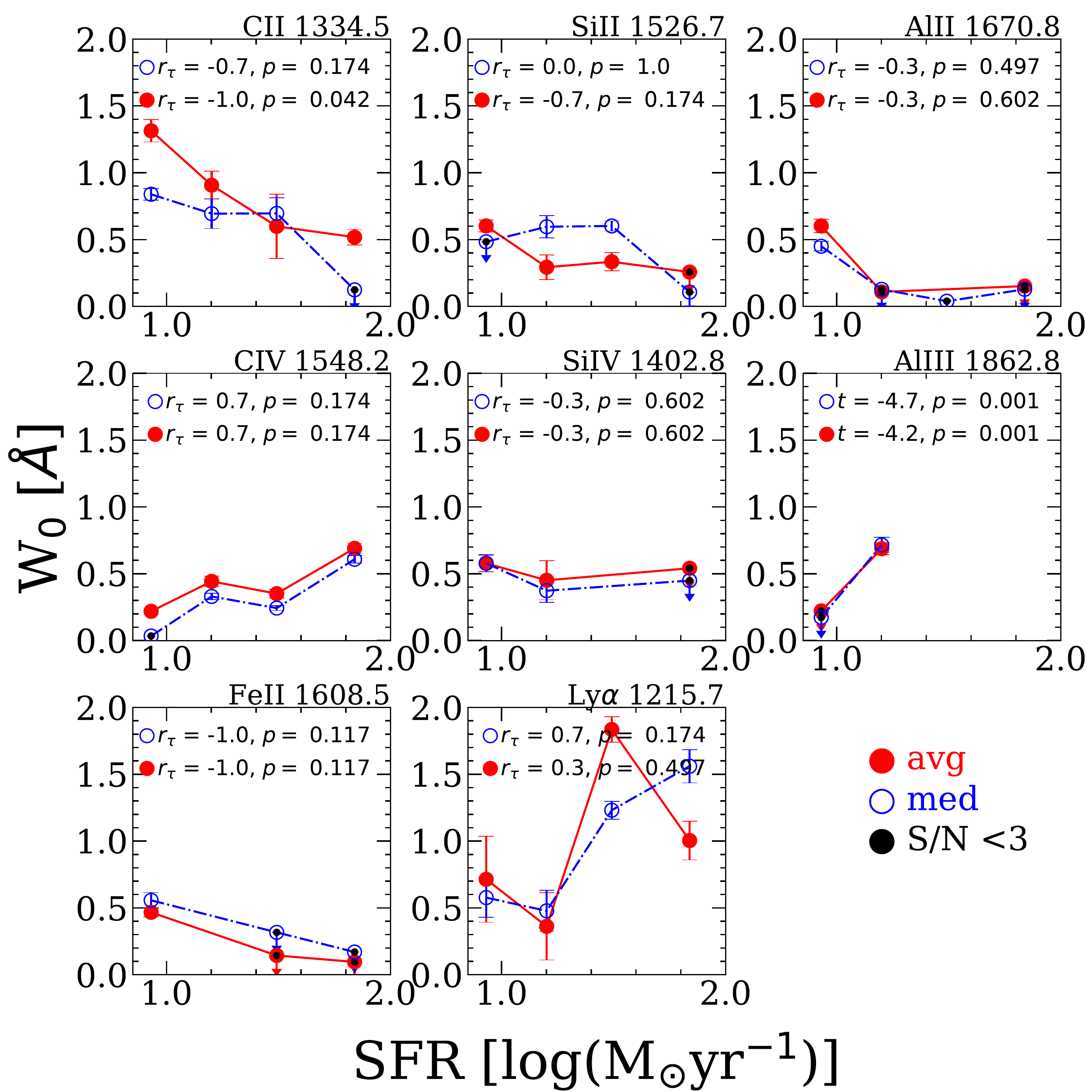} \\
 \includegraphics[width=85mm]{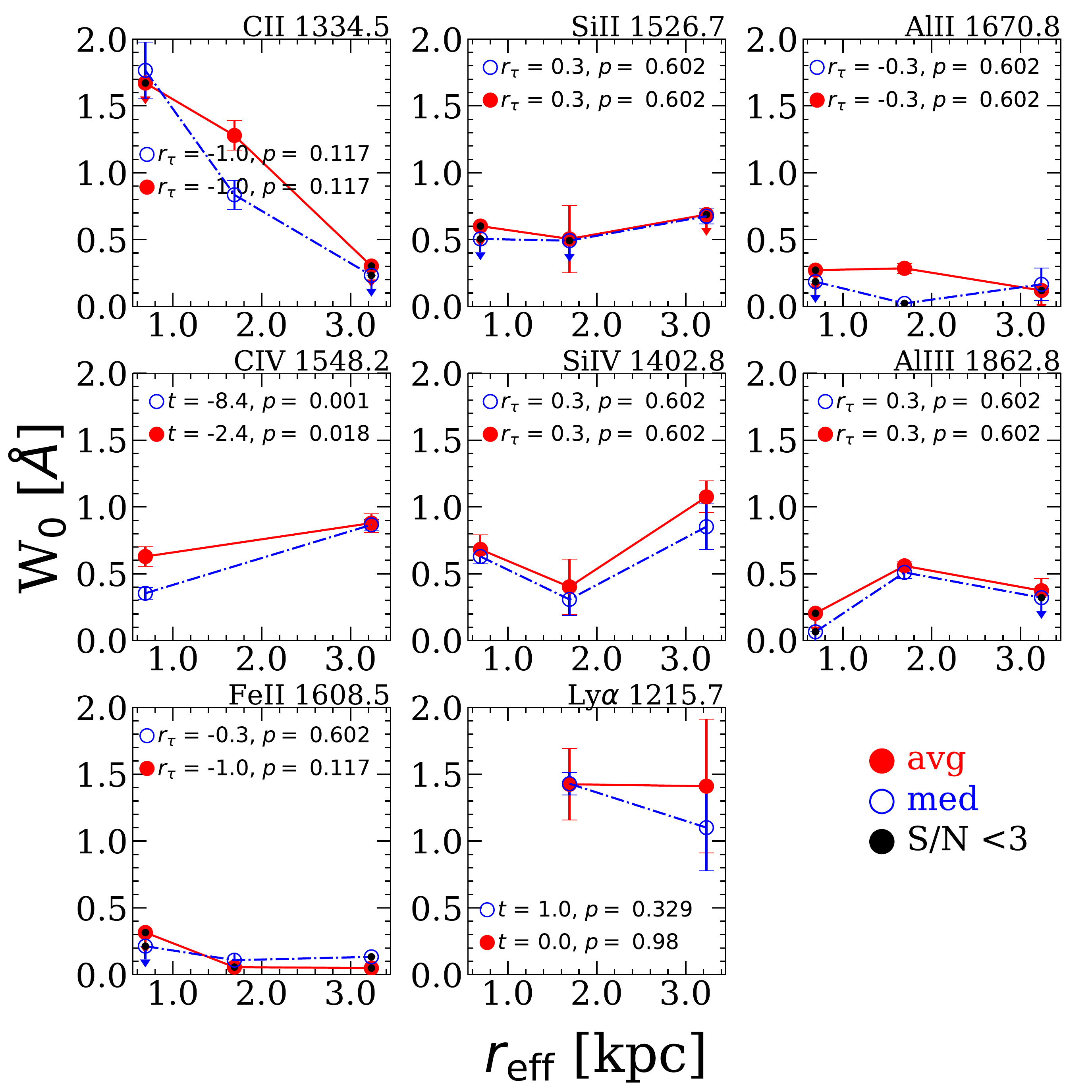} &
 \includegraphics[width=85mm]{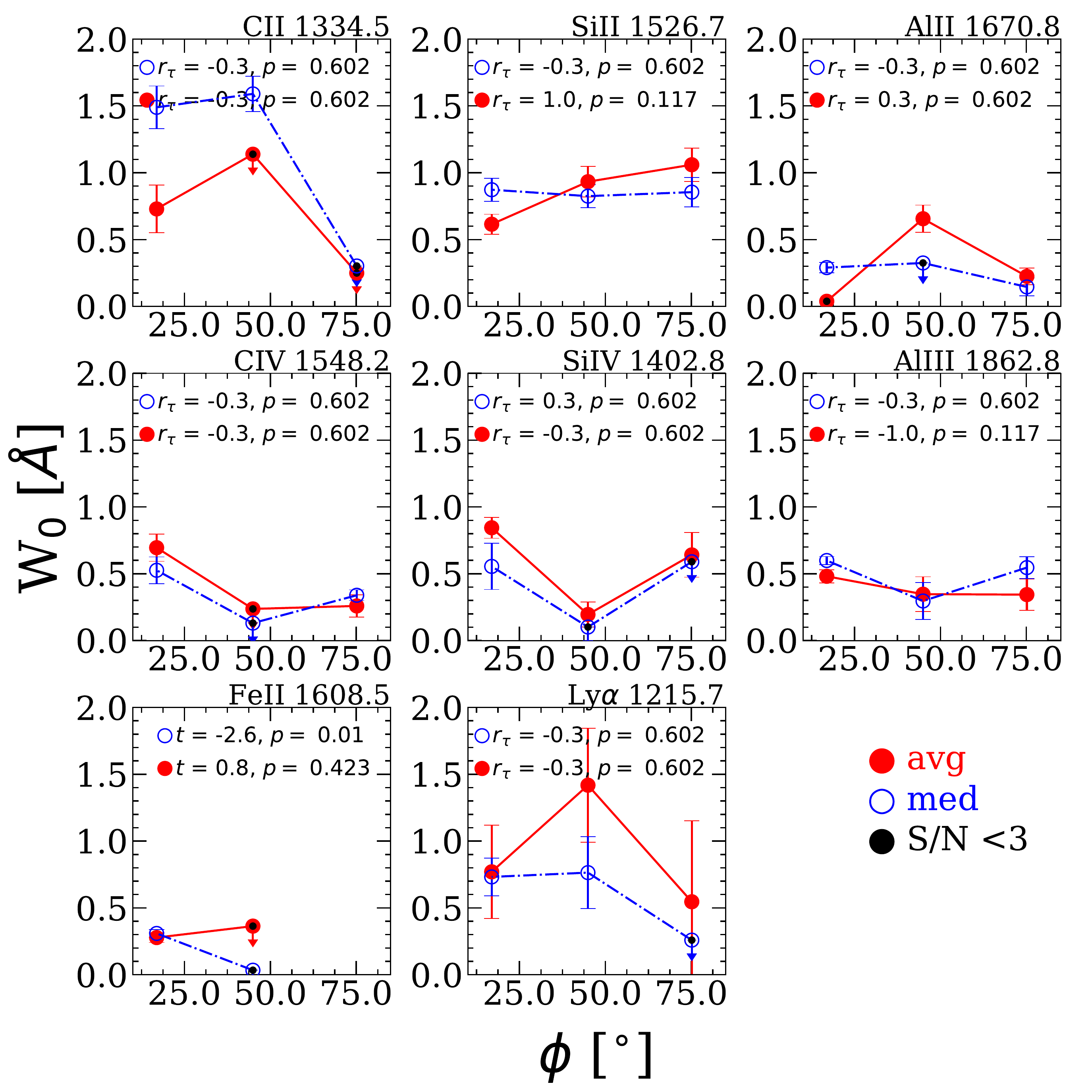} \\
\end{tabular}
\caption{Rest equivalent width ($W_{0}$) of the absorption lines as a function
 of the \textit{foreground} galaxy's stellar mass (log[M$_\bigstar$/M$_
 {\odot}$]), SFR (log[M$_{\odot}$yr$^{-1}$]), effective radius ($r_{\rm
 {eff}}$), and azimuthal angle($\phi$). The $W_{0}$ measurements come from
 composite spectra considering galaxy pairs at all projected
 distances ($b < 23 \arcsec$) and split by the corresponding galaxy property.
 Stellar mass and SFR $W_{0}$ were obtained from composite spectra
 considering 238 \textit{bg} galaxies, while $r_{\rm{eff}}$ and $\phi$ come
 from composite spectra considering 97 \textit{bg} galaxies (see Sect. \ref
 {sec:Data} and Table \ref{Tbl:Samples}). Average and median $W_{0}$ are
 shown in solid red and open blue symbols; solid black symbols
 correspond to upper limits with S/N <3. The error bars correspond to 1$\sigma$
 confidence intervals for average or median values based on a bootstrap
 analysis. The panels include the results from the Kendall-Tau correlation
 test: the correlation coefficients $r_{\tau}$ and their corresponding
 $p$-value (the probability of no correlation). The panels of absorption
 lines with detections in only two subsamples also show the results from a
 Student's t-test:  the difference between a pair of mean values given by the
 $t$ coefficient and their corresponding $p$-value (the probability of
 significant difference amongst means).}
\label{Fig:EW-Pfl-Bg-PRP-ALL}
\end{figure*}

To explore the dependence of low- and high-ionization state absorptions with
physical and morphological properties (i.e. star formation rate, stellar
mass, galaxy effective radius, and azimuthal angle) produced in the CGM  of
our high-redshift star-forming galaxy sample, we considered galaxy pairs at
all projected distances ($b < 23 \arcsec$) split by the corresponding galaxy
property. A summary of the statistical properties of the galaxy pair
subsamples divided by these properties can be found in  Appendix \ref
{sec:AppendixTbl}. We note that the  Fe\,{\sc ii}, Al\,{\sc iii}, and Al\,
{\sc ii} lines were excluded from our following analyses (see Sect.  \ref
{sec:EW-Rad}). Figure \ref{Fig:EW-Pfl-Bg-PRP-ALL} (upper panels) shows the
equivalent width of the absorption features as a function of stellar mass and
SFR for LIS (C\,{\sc ii}, Si\,{\sc ii}), HIS (Si\,{\sc iv}, C\,{\sc iv}), and
Ly$\upalpha$. We find that galaxies with high stellar masses (log[M$_
{\bigstar}$/M$_{\odot}$]>10.30) and high star formation rates (log[SFR/(M$_
{\odot}$yr$^{-1}$)]>1.93) show C\,{\sc iv} (HIS) metal absorption with larger
equivalent widths. On the other hand, galaxies with low stellar masses (log
[M$_{\bigstar}$/M$_{\odot}$]<9.26) and low SFRs (log[SFR/(M$_{\odot}$yr$^
{-1})]<0.9$) show C\,{\sc ii} (LIS) metal absorptions with stronger
equivalent widths. Nevertheless,  Si\,{\sc ii} and  Si\,{\sc iv} do not show
a similar trend. This might be caused by the fact that we do not have Si\,
{\sc ii} and  Si\,{\sc iv} detections with S/N>3 in all of our SFR and
stellar mass bins.

To quantify the robustness of the trends highlighted in Figure~\ref
{Fig:EW-Pfl-Bg-PRP-ALL}, we ran a Kendall-Tau rank test for the absorption
lines that are detected in at least three bins. We find that SFR and stellar
mass are anticorrelated with C\,{\sc ii}, but they correlate positively with
C\,{\sc iv}. This would imply that C\,{\sc ii} is located mostly in galaxies
with low stellar mass(log[M$_{\bigstar}$/M$_{\odot}$]<9.26) and low SFR (log
[SFR/(M$_{\odot}$yr$^{-1})]<0.9$), while C\,{\sc iv} is detected in galaxies
with high stellar mass (log[M$_{\star}$/M$_{\odot}$]>10.2) and high SFR (log
[SFR/(M$_{\odot}$yr$^{-1})]>1.5$). The slopes of the $W_{0}$ relationships
with stellar mass and SFR for C\,{\sc ii} and C\,{\sc iv} are 0.79 $\pm$
0.13, 0.61 $\pm$ 0.05 and 0.73 $\pm$ 0.08, 0.55 $\pm$ 0.04, which are
significantly different at $\geq5\sigma$ level. We do not find any robust
correlation between SFR and stellar mass with Si\,{\sc ii}, Si\,{\sc iv},
Fei\,{\sc ii}, Al\,{\sc ii}, Al\,{\sc iii}, or Ly$_{\upalpha}$, yielding a
probability of no correlation above 60\% in all cases. We also explored the
combined effect of impact parameter with stellar mass and SFR; however, no
clear conclusion was reached due to the low number statistics, which resulted
in composite spectra with low S/N line detection.

\subsection{Morphological dependence}\label{sec:EW-Mrph}

The strength of the observed metal absorption signatures in the CGM can be
explored as a function of the azimuthal angle $\phi$ between the l.o.s. and
the projected major and minor axes of the
\textit{foreground} galaxy. Here we define the azimuthal angle ($\phi$) as the
projected angle between the \textit{background} galaxy l.o.s. and the
projected minor axis of the \textit{foreground} galaxy. Small azimuthal
angles ($\phi\sim0^{\circ}$) refer to l.o.s. passing along the projected
minor axis of the \textit{foreground} galaxy, while large azimuthal angles
($\phi=90^{\circ}$) refer to l.o.s. passing along the projected major axis
of the \textit{foreground} galaxy. The effective radius ($r_{\rm
{eff, circ}}=r_{\rm{eff}}\sqrt{q}$ 
\footnote{$q$ is the axis ratio (b/a) of the elliptical isophotes that best
fit the galaxy.}) was obtained by fitting a single S\'ersic profile with no
constraints on the parameters, and then was circularized using the galaxy
ellipticity \citep{Ribeiro16}. Alongside the effective radius, {\sc Galfit}
provides other structural parameters: S\'ersic index (n), the axis ratio of
the ellipse (b/a), and the position angle, $\theta_{\rm{PA}}$. Out of the
238 galaxies evaluated in this paper, only 97 are in the COSMOS field and
covered by HST imaging, and therefore have morphological parameters. To
inspect the dependency of the LIS-HIS absorption strengths with the
azimuthal angle($\phi$), we split our sample into three $\phi$ bins:[0-30]$^
{\circ}$,[30-60]$^{\circ}$, and [60-90]$^{\circ}$. Figure~\ref
{Fig:EW-Pfl-Bg-PRP-ALL}(lower panels) reports the equivalent widths of
different spectral features in bins of $r_{\rm{eff}}$ and $\phi$.

To check and assess the robustness of possible correlations between the
morphological parameters and the equivalent widths measured in the composite
spectra for LIS, HIS, and Ly$_{\upalpha}$, we ran a Kendall-Tau rank test
between $W_{0}$ and $\phi$ and $r_{\rm{eff}}$. For lines detected in only two
bins  the correlation cannot be assessed; therefore, we estimated the
significance of the difference between their absorption strength ($W_
{0}$). To do this  we applied a Student's t-test to determine the probability
that the $W_{0}$ variations between the low and high $r_{\rm{eff}}$ and
$\phi$ populations are not statistically significant. According to the
Student's statistics ($t$) a large $p$-value indicates to a high probability
that the null hypothesis is correct (the two samples are not statistically
significant); a small $p$-value suggests that the difference is significant.
The  Kendall-Tau and Student's t-test results are both given in Figure \ref
{Fig:EW-Pfl-Bg-PRP-ALL} in the corresponding panels. 

For the effective radius ($r_{\rm{eff}}$) of the galaxy, the Kendall-Tau test
shows a mild anticorrelation with C\,{\sc ii} ($p<0.117$), while it does not
show any significant correlation with Si\,{\sc ii} and Si\,{\sc iv}
($p>$0.602). We find a significant difference ($p<0.01$) in the $W_
{0}$ variations between small and large $r_{\rm{eff}}$ as shown by the
Student's t-test. However, no correlation is found for Ly$\upalpha$. This
suggests that C\,{\sc iv} gas is usually located in the CGM of larger
galaxies, while C\,{\sc ii} gas is located in the CGM  of smaller galaxies.
Concerning the azimuthal angle, we do not find any significant correlation
with $\phi$ in any of the spectral lines inspected in agreement with what has
been reported by \citep{Chen21}, who inspected the azimuthal dependence of
Ly${\upalpha}$ emission of 59 star-forming galaxies ($z_{med}\sim 2.3$). This
is opposed to the low-redshift scenario where galaxies show high-velocity
biconical outflows oriented along the minor-axis and accreting material along
the major-axis. This result could be linked to the metal distribution along
the disc of star-forming galaxies as metal absorption systems are often
compact and poorly mixed, where cool low-ionization metal absorbers have
typical sizes of $\sim 1$ kpc \citep{Faucher-Giguere17}, while
high-ionization gas seen in absorption arises in multiple extended structures
spread over $\sim$100 kpc \citep{Churchill15,Peeples19}.

\subsection{C\,{\sc ii}/C\,{\sc iv} line ratio}\label{sec:LR}

\begin{figure*}
   \centering
   \includegraphics[width=18cm]{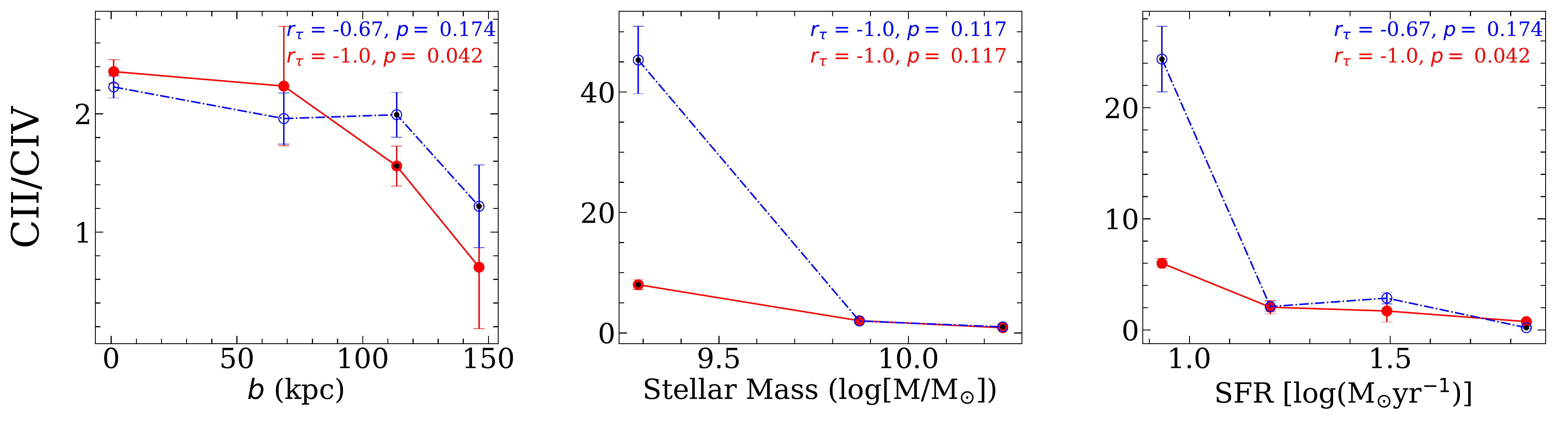}
   \caption{C{\sc ii}/C{\sc iv} $W_{0}$ line ratio as a function of the impact
    parameter $b$, and the \textit{foreground} galaxy's stellar mass and SFR.
    C{\sc ii}/C{\sc iv} $W_{0}$ values were obtained considering the  same set
    of \textit{fg-bg} galaxy pairs. Average and median $W_{0}$ are shown in
    solid red and open blue symbols; the  solid black symbols
    correspond to upper limits with S/N <3. The error bars correspond to
    1$\sigma$ confidence intervals for average or median values based on a
    bootstrap analysis.}
    \label{Fig:EW-C2C4LR}
    \end{figure*}

A further inspection of the differences between LIS and HIS on the different
galaxy parameters come from the C\,{\sc ii}/C\,{\sc iv} equivalent width line
ratio. We focus on carbon ions as they provide stronger absorptions with
higher S/N compared to aluminium or silicate ions. Figure~\ref
{Fig:EW-C2C4LR} shows the  C\,{\sc ii}/C\,{\sc iv} line ratio as a function
of impact parameter, stellar mass, and star formation rate. The  C\,
{\sc ii}/C\,{\sc iv} line ratio appears to  anticorrelate with impact
parameter($b$), stellar mass, and star formation rate. Similarly to the
analyses for individual line absorptions, we ran a Kendall-Tau to assess
possible correlations  for C\,{\sc ii}/C\,{\sc iv} line ratio and included it
in all panels of Figure~\ref{Fig:EW-C2C4LR}. We confirm that the
anticorrelations between C\,{\sc ii}/C\,{\sc iv} ratio and impact parameter,
stellar mass, and star formation rate are statistically robust. This suggests
that C\,{\sc ii} is more important than C\,{\sc iv} in the inner regions of
these star-forming galaxies, while the opposite occurs in the outskirts at
large separations. 

On the other hand, star-forming galaxies and low stellar mass with low star
formation rates show a higher C{\sc ii}/C{\sc iv} line ratio compared with
galaxies that have high stellar mass and high star formation rates. Our
results suggest that galaxies with higher star formation rates and large
stellar masses are capable of sweeping out the highly ionized gas (traced by
C\,{\sc iv}) farther away from the galaxy compared with less actively
star-forming and less massive galaxies. Another possible explanation is that
more active and massive galaxies have stronger ionizing fluxes, able to
ionize gas at larger distances compared with less massive and less active
galaxies. Recent studies have shown that fast and energetic outflows can push
material away from the central regions in star-forming galaxies, more
effectively in galaxies with high SFR (with a weaker dependence on stellar
mass; 
\citealt{Heckman15,Cicone16,Trainor15,Feltre20}.) The presence of an AGN can
make this effect even more dramatic, as shown by an enhanced gas budget in
{\sc H\, i} and low-ionization gas in high-redshift ($z>2$) galaxies \citep
{Prochaska14}, suggesting higher accretion rates or a net gain of cold gas
in the CGM in these star-forming galaxies \citep{Faucher16}. However, the
physical mechanism through which the AGN removes and/or heats the gas and
suppresses accretion is not clear \citep{Tumlinson17}. Determining the
nature of the source responsible for ionizing the gas in the CGM is crucial
to improving our understanding of the multi-phase CGM, but is beyond the
scope of this paper. 

\subsection{Ly$\upalpha$ emission}\label{sec:Ly-Emm}

\begin{figure}
\centering
\includegraphics[width=8.5cm]{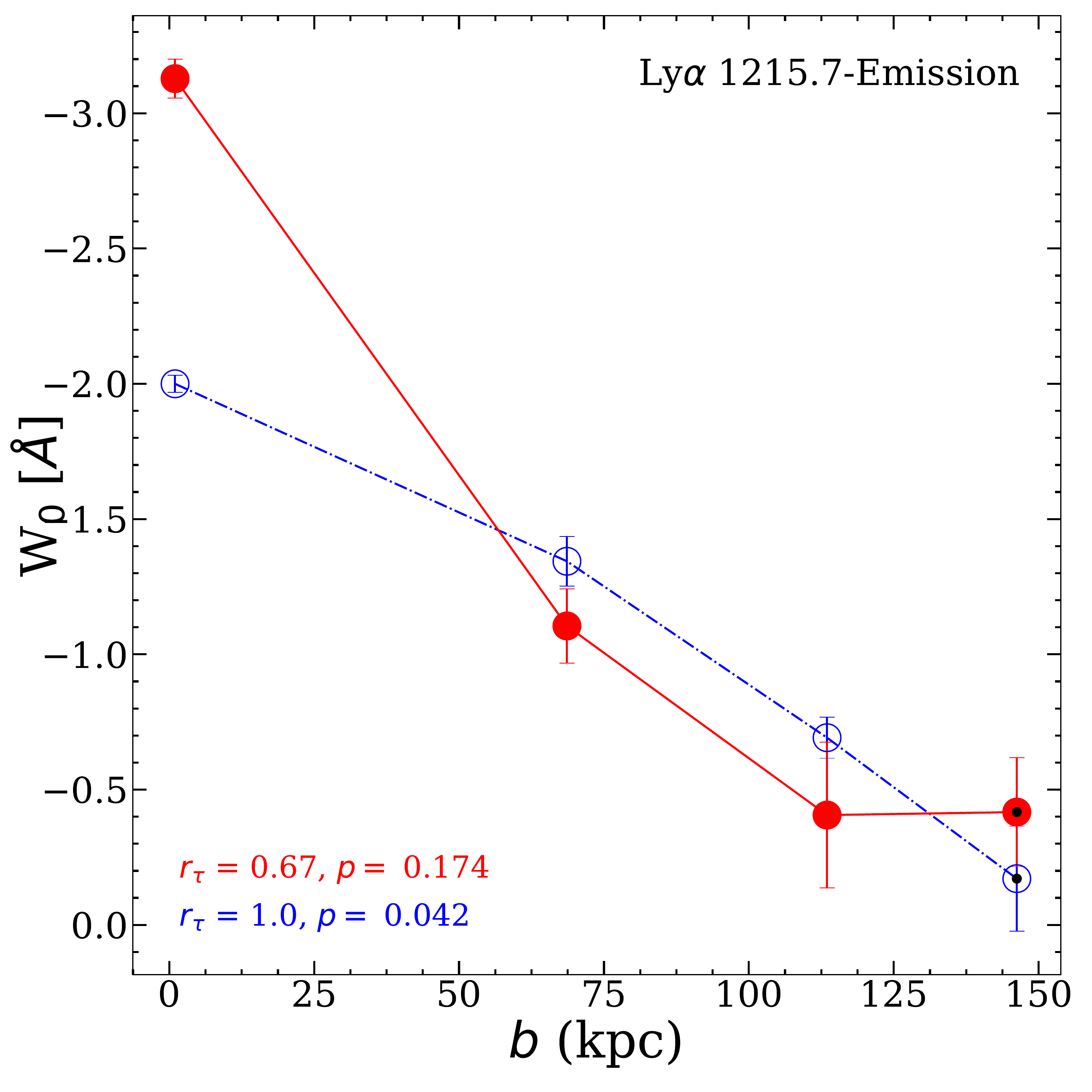}
\caption{Equivalent width ($W_{0}$) for the Ly${\upalpha}$ emission as a
 function of the impact parameter ($b$) obtained from the average
 (solid red line) and median (dot-dashed blue line) composite spectra. The error
 bars correspond to 1$\sigma$ confidence intervals 
 for average or median values based on a bootstrap analysis.}
\label{Fig:EW-Lya-Emm}
\end{figure}

As we already noted in section~\ref{sec:EW-Rad} (Figure~\ref
{Fig:Lns-Bg-Sep-4}), the Ly$\upalpha$ emission component is clearly detected
in our composite spectra. Figure \ref{Fig:EW-Lya-Emm} shows the Ly$\upalpha_
{\rm{em}}$ rest equivalent width as a function of the impact parameter. We
find that Ly$\upalpha_{\rm{em}}$ decreases as a function of the impact
parameter  $b$ similarly to what \cite{Chen21} reported from 2D Ly$\upalpha_
{\rm{em}}$ maps. We also explore the Ly$\upalpha_{\rm{em}}$ equivalent width
strength as a function of the galaxy stellar mass, star formation rate, and
azimuthal angle. However, we do not find any significant correlation in these
cases. This might be caused by the fact that at $z\sim$2 the Ly$\upalpha_{\rm
{em}}$--stellar mass relation is weaker compared with galaxies at higher
redshifts \citep{Du18}. Another possibility is that our Ly$\upalpha_{\rm
{em}}$ measurements come from composite spectra, hence  from star-forming
galaxies with and without direct Ly$\upalpha_{\rm{em}}$ detections
(EW$\leq0\AA$, EW$>0\AA$ and EW$\geq20\AA$ \citealt
{LeFevre15,Hathi16}), covering a wide range of stellar masses and star
formation rates that may have diluted a real Ly$\upalpha_{\rm{em}}$ signal
\citep{Steidel11}.

\section{Discussion}\label{sec:Dis}

Analyses of low- and high-ionization state absorption features and their
dependence on stellar mass, star formation rate, and galaxy inclination have
been widely explored at $\lesssim0.5$ (see \citealt
{Tumlinson17,Kacprzak17}). At high redshift, however, the studies of the CGM
are limited. \cite{Steidel10} used 512 close ($<15\arcsec=124$kpc) angular
pairs of $z\sim$2--3 ($z_{bg}\sim$2.3) galaxies to map the cool and diffuse
gas around galaxies. They found strong evidence suggesting the presence of
superwind outflows, and proposed a simple model of the gas in the CGM, where
cool gas is distributed symmetrically around every galaxy, accelerating
radially outwards with the outflow velocity increasing with radius. Later,
\cite{Turner14} and \cite{Lau16} reported detections that suggest that in
galaxies at $\langle z \rangle$\,=\,2.4 the CGM extends at least up to
180\,kpc. Moreover, the mass of metals found within the halo is substantial
and equivalent to $\sim$25\% of the metal mass within the interstellar
medium \citep{Rudie19}. This is, however, a lower percentage than what has
been reported from studies in low-redshift galaxies \citep{Werk14}, which
suggests a considerable redistribution of the metal content of galaxies in
an inside-out fashion over the last $\sim$8.5 Gyr. We note that eight of the
nine resonance Si\,{\sc ii} transitions fall within the
[900-1900]$\AA$ wavelength range considered here: 
$\uplambda 989 \AA$,
$\uplambda 1020 \AA$, $\uplambda\uplambda 1190, 1193 \AA$, 
$\uplambda 1260 \AA$, $\uplambda 1304 \AA$ $\uplambda 1526 \AA$, and 
$\uplambda 1808 \AA$ \citep {Shull81}. Lines located bluewards 
of Ly$_{\upalpha}$ are not included in our analyses as the result of the low
coverage of our composite spectra at this wavelength range and because UV
continuum level bluewards of Ly$_{\upalpha}$ is particularly sensitive to the
dust content, metallicity, and the age of the stellar population \citep
{Trainor15} and their strength could be affected by IGM absorptions \citep
{Shapley03}. The Si\,{\sc ii} ($\uplambda1304 \AA$) component is blended with
the   O\,{\sc i} ($\uplambda 1302 \AA$) absorption line and is close to
excited fine-structure emission transitions that could lower the  $W_
{0}$ measurements \citep{Trainor15}. Finally,  Si\,{\sc ii}
($\uplambda 1526 \AA$) is the least resonance component affected by
blends \citep{Shapley03} and is the strongest Si\,{\sc ii} line amongst their
counterparts. For this work we chose $\uplambda 1526 \AA$ for our Si\,
{\sc ii} $W_{0}$ computations as this component is free of blending effects
caused by  the low spectral resolution of our dataset. However, as noted
by \cite{Jones18} Si\,{\sc ii} transitions probe diverse optical depths, and
hence the most accurate way to compute Si\,{\sc ii} absorption strength is to
consider the contributions of all the Si\,{\sc ii} resonance components
($\uplambda\uplambda\uplambda 1260, 1304, 1526 \AA$) within the analysed
range, by averaging their strengths and deblending them from nearby spectral
features that could affect their individual $W_{0}$ measurements.
Nevertheless, different studies have adopted diverse strategies to compute
the LIS (including Si\,{\sc ii} resonance components) $W_{0}$ strength. For
example, \cite{Jones18} averaged Si\,{\sc ii}
($\uplambda\uplambda\uplambda 1260, 1304, 1526\AA$), O\,{\sc i}
($\uplambda 1302$), C\,{\sc ii} ($\uplambda 1334 \AA$), Al\,{\sc ii}
($\uplambda 1670 \AA$), Fe\,{\sc ii} ($\uplambda 2382 \AA$), and Mg\,
{\sc ii} ($\uplambda\uplambda 2796, 2803 \AA$) components; 
\cite{Ranjan22} considered the average of the eight Si\,{\sc ii} resonance
components to estimate the $W_{0}$ Si\,{\sc ii}  absorption in high-z damped
Lyman-$\upalpha$ (DLA) systems; \cite{Du16} averaged 
Si\,{\sc ii} ($\uplambda 1526 \AA$), Al\,{\sc ii} ($\uplambda 1670 \AA$), 
Ni\,{\sc ii} ($\uplambda\uplambda1741, 1751 \AA$), 
Si\,{\sc ii} ($\uplambda 1808 \AA$), and Fe\,
{\sc ii} ($\uplambda 2382 \AA$)
absorption lines to compute an overall LIS average absorption estimation,
while later  
\cite{Du18} considered Si\,{\sc ii} ($\uplambda 1260 \AA$), O\,{\sc i}--Si\,
{\sc ii} ($\uplambda\uplambda1302, 1304 \AA$), C\,{\sc ii}
($\uplambda 1334 \AA$), and Si\,{\sc ii} ($\uplambda 1526 \AA$) for similar
computations. The differences in the number of Si\,{\sc ii} resonance
components considered for $W_{0}$ computations could be the reason why
similar Si\,{\sc ii} trends to those of C\,{\sc ii} with SFR and stellar
mass  shown in Fig. \ref{Fig:EW-Pfl-Bg-PRP-ALL} are not appreciated. As
shown by \cite{Jones18} different low ion transitions (Si\,{\sc ii} , Fe\,
{\sc ii}, and Ni\,{\sc ii}) show non-uniform covering fractions in the CGM
of star-forming galaxies at high redshift ($z\simeq2-3$).

In this work we  used a sample of 238 galaxy close pairs to probe the CGM
around star-forming galaxies at $z\sim2.6$. Our results show an
anticorrelation between the equivalent width of the absorption lines and the
impact parameter ($b$) for this high-redshift star-forming galaxy sample and
are consistent with previous results at lower redshift. Here we detect
Ly$\upalpha$--LIS and HIS absorption at distances up to 172\,kpc and
146\,kpc. Our low-ionization state (C\,{\sc ii}, Si\,{\sc ii}), and
high-ionization state (C\,{\sc iv} Si\,{\sc iv}) absorption line detections
shape an upper envelope of the equivalent width distribution coming from
studies at low redshift (see Figure \ref{Fig:EW-Pfl-Bg-sep-4}). This is
further illustrated by the differences in the slopes of the $W_{0}$ radial
profiles between low- and high-ionization state absorptions of the same ion
species, where Si\,{\sc ii}--Si\,{\sc iv} and C\,{\sc ii}--C\,{\sc iv}
absorption line pairs are different at a $\geq5\sigma$ level. Moreover, when
compared with galaxies at $z\sim2.3$ \citep{Steidel10} our results show a
significant difference in their slopes at a 5$\sigma$ level for Si\,{\sc ii}
and a 3$\sigma$ level for C\,{\sc ii} and Al\,{\sc ii} absorption lines.
These discrepancies between LIS and HIS suggest that, within the CGM, cold
(T$<10^{4.5}$K) and dense gas is more extended in galaxies at $z\sim2.6$
compared with galaxies at $z\sim2.3$ and lower redshifts, as probed by Si\,
{\sc ii}, C\,{\sc ii}, Si\,{\sc iv}, and  C\,{\sc iv}. These results suggests
a potential redshift evolution for the CGM gas content that produces these
absorptions. As higher covering factors of neutral gas lead to stronger
absorption lines \citep{Du18}, this suggests a higher content of neutral gas
in high-redshift galaxies compared with low-redshift galaxies. This
concentration of neutral gas could represent a reservoir of gas that can be
later accreted onto the galaxies to fuel future star formation \citep
{Hafen19,Hafen20}.

We note that it is possible to increase the maximum angular projected
separation ($\geq23\arcsec$) allowing us to probe larger distances closer to
the CGM and IGM boundary, following \cite{Chen20}. However, we  limited our
analyses to 23$\arcsec\sim$170kpc and galaxy spectra with highly reliable
redshifts to guarantee detection with good S/N. Spectra with unreliable
redshift measurements could wash out the expected metal absorptions and/or
produce spurious features in the final composite spectra. This is further
supported by the observed metallicity dependence with impact parameter, where
LIS are preferably located in the inner CGM (close to the galaxy; \citealt
{Lau16}), while HIS (e.g. O\,{\sc vi} are expected to dominate the CGM at
larger separations \citealt{Shen13}) making it difficult to detect weak LIS
absorptions at larger separations given their expected $W_{0}$.

\subsection{LIS and HIS absorption dependencies.}

At $z \lesssim1$, \cite{Churchill00b} explored a sample of 45 Mg\,{\sc ii}
absorption systems in high-resolution QSO spectra, and studied their spectra,
together with the C\,{\sc iv} and Fe\,{\sc ii} absorption profiles,
suggesting that the gas responsible for Mg\,{\sc ii} and C\,{\sc iv}
absorptions arises from gas in different phases (i.e. gas producing HIS
absorption features have typical temperatures T>10$^{5}$K \citealt
{Tumlinson17}). Moreover, they observed an evolution of the absorbing gas
that is consistent with scenarios of galaxy evolution in which mergers and
accretion of `protogalactic clumps' provide gas reservoirs responsible for
the elevated star formation activity at high redshift, while at intermediate
and lower redshifts ($z\leq$1), the balance of high- and low-ionization state
CGM gas may be related to the presence of star-forming regions in the host
galaxy \citep{Churchill00b}. Later on, \cite{Zibetti07} found that
early-type (quiescent, red) galaxies ($0.37 < z < 1$) are associated with
weaker  Mg\,{\sc ii} absorptions, while stronger systems are located in
late-type (star-forming, blue) galaxies. These results provided evidence that
the SFR correlates with Mg\,{\sc ii} absorption strength caused by the
presence of outflows from star-forming or bursting galaxies.  This scenario
is also supported by subsequent studies that demonstrated that SFR correlates
with outflow velocity deduced from the Mg\,{\sc ii} absorption strength 
\citep{Weiner09, Martin12,Tremonti07,Rubin10}, and by the correlation between
the stellar mass and Ly$\upalpha$ absorption strength \citep
{Bordoloi18,Wide21}.

Our detection of LIS and HIS absorption lines suggests that at $\langle
z \rangle \sim 2.6$ the C\,{\sc ii} and C\,{\sc iv} features are correlated
with star formation rate and stellar mass. C\,{\sc ii} is stronger in
galaxies with low star formation rates and low stellar masses, while C\,
{\sc iv} is stronger in galaxies with high SFR and high stellar masses. These
results are consistent with what has been reported for low-redshift galaxies
in scenarios where LIS and HIS absorbers have different spatial
distributions. Low-ionization state metal absorbers probe the gas that is
less ionized than high-ionization state metal absorbers \citep{Nagao06}, and
thus they are expected to be located in gas regions with different density
conditions \citep{Burchett16}. Moreover, studies of low-redshift galaxies
($z\leq0.5$) have shown that as a consequence of the interaction between a
starburst-driven wind and the preexisting CGM at radii as large as 200\,kpc,
the CGM around star-forming galaxies with high star formation rates differs
systematically compared to galaxies with lower SFRs, as probed  by the
Ly$\upalpha$, Mg\,{\sc ii}, Si\,{\sc ii}, C\,{\sc iv}, and O\,{\sc vi}
absorption lines \citep{Tumlinson11a,Borthakur13,Lan14,Heckman17}. However,
as suggested by 
\cite{Cicone16} and \cite{Gatkine22}, their correlation with stellar mass and
star formation rate might in fact  be a consequence of a main-sequence
offset, rather than simply a correlation with star formation rate or stellar
mass,  and because LIS--HIS covering factors can be affecetd by
environmental processes  \citep{Dutta21}. Hence, caution should be taken
when interpreting LIS--HIS correlations with SFR and/or stellar mass.

The origin of the reservoir of cold neutral gas around low stellar mass
galaxies and low SFRs at these distances($\langle b \rangle\approx 120$kpc;
see Appendix \ref{sec:AppendixTbl}) is uncertain. One explanation comes from
a soft radiation field unable to ionize HIS (i.e. C\,{\sc iv}, Si\,{\sc iv})
as the production of C\,{\sc iv} and Si\,{\sc iv} absorption features in the
CGM requires photons with energies >45 eV associated with hard ionizing
radiation fields from massive stars, AGN, and radiative shocks \citep
{Feltre20,Trainor15}. As shown by \cite{Gatkine22}, HIS are more affected by
high-velocity outflows; they show a strong correlation with SFR and are
ubiquitous in high-SFR systems. Another possible explanation comes from the
galactic fountain scenario, predicted by cosmological galaxy formation
simulations \citep{Oppenheimer10,Vogelsberger13},  where the presence of cold
low-ionization gas results from metal-rich gas ejected in previous star
formation episodes that fall back to the disc \citep
{Fraternali06,Hobbs15}. Moreover, we know that stellar winds are dominant in
the inner region of the CGM <60kpc \citep{Chen20}, and depending on their
velocity, stellar winds in low-mass galaxies might be able to expel material
or be reaccreted as recycled material \citep{SanchezA17}. However, if the
velocity outflow is insufficient to eject material out into the IGM, the
recycle timescales in low-mass galaxies where gas returns to the galaxy can
be as small as those in high-mass systems 
\citep{vandeVoort17}.  \cite{Martin12} detected Fe\,{\sc ii} Doppler shifts
and  $V_{max}$--Mg\,{\sc ii} values that suggest that outflows reach the
circumgalactic medium with Mg\,{\sc ii} absorption at blueshifts as high as
700 km s$^{-1}$, reaching 70 kpc in 100 Myr, a short enough time for the
host galaxy to sustain SFR, even if the SFR declines due to an outflow.
Nevertheless, our measurements do not allow us to break down the multiple
components of the absorption features in our \textit{foreground} galaxy
(down-the-barrel) composite spectra and to detect blueshifted offsets that
could be related to high-velocity outflows. We note that it is possible to
combine down-the-barrel and QSO--galaxy and/or galaxy--galaxy pairs to
unambiguously detect blueshifted absorptions relative to the galaxy systemic
velocity, and quantify independently the main properties of the detected
outflow \citep{Kacprzak14,Bouche16,Lehner17}.

Figure \ref{Fig:EW-Pfl-Bg-PRP-ALL} (lower panels) shows the dependence of line
absorption strength on the galaxy's effective radius ($r_{\rm{eff}}$) and
azimuthal angle ($\phi$). For the dependence of the absorption strength ($W_
{0}$) on the size of the galaxy given by the effective radius($r_{\rm
{eff}}$), we find that LIS absorptions are stronger in smaller galaxies
compared to HIS absorptions, which are shown to be stronger in larger
galaxies, in agreement with \cite{Rudie19}, who reported that LIS absorption
occurs spatially closer to the galaxy, and gas producing HIS absorptions can
be located well beyond the virial radius. Regarding the dependence on
azimuthal angle, we do not find any significant correlation with $\phi$ for
either LIS or HIS. However, these results are based on the fraction of
galaxies for which we have morphological measurements ($\sim 40\%$ of the
parent sample). Thus, these results have to be confirmed by a similar
morphological analysis using larger samples at these high redshifts, as
galaxy orientation plays a major role in the measured strength of LIS and
HIS. This has been shown at low redshift from studies using analyses of
down-the-barrel \citep{Rubin14,Bordoloi14b} and QSO--galaxy pairs \citep
{Kacprzak11b,Bordoloi11}. Moreover, at z $\gtrsim2$ star-forming galaxies
have clumpy morphologies \citep{Foster11}, with galactic winds that are
mainly driven by outflows from prominent star-forming clumps \citep
{Genzel11}, and have not yet formed  a stable disc (or any disc at all)
capable of collimating galactic winds into bipolar outflows \citep
{FaucherG17}, indicating that the minor--major axis dichotomy associated with
rotation present at low redshift (z<1) is not broadly applicable at
z$\sim$2-3 
\citep{Law12,Nelson19,Price20}. Therefore, it is not obvious that at high
redshift ($z\gtrsim2$) the azimuthal angle can be determined and, if it can,
whether it has any physical meaning.

The dependence of LIS and HIS absorption strength on the azimuthal angle has
been explored in the literature to prove or disprove theories on how galaxies
accrete gas from the intergalactic medium. Inflowing material along filaments
in the IGM is expected to be the largest source of accretion \citep
{Keres05}. Simulations have shown that accretion of metal-poor gas inflows
occurs along the major axis of galaxies, where outflows that are preferably
located along the semi-minor axis form bipolar outflows \citep{Putman17}, and
gas being accreted onto galaxies that can later   trigger star formation has
been supported by  evidence collected from multiple CGM studies at low
redshift \citep
{Kacprzak10,Kacprzak12a,Kacprzak15a,Kacprzak19,Lan18,Martin19a,Nielsen15,
Rubin18b,Zhu13a,Tumlinson11a,Bordoloi14a,Lan14}. Recent cosmological
hydrodynamical simulations examine the physical properties of the gas located
in the CGM of star-forming galaxies as a function of angular orientation
by \citet{Peroux20b}. They found that the CGM varies strongly with impact
parameter, stellar mass, and redshift. Moreover, they suggest that the inflow
rate of gas is more substantial along the galaxy major axis, while the
outflow is strongest along the minor axis. 

The correlations between  C\,{\sc ii} and C\,{\sc iv} with impact parameter,
star formation rate, and stellar mass scenario of a multi-phase CGM(where LIS
absorptions are produced by denser gas located closer to the galaxy, whether
this dense neutral gas component is part of an outflow or material falling
back to the galaxy taking part in a galactic fountain that could eventually
be funnelled to the galaxy to sustain star formation activity;  \citealt
{Keres05}) is not clear and cannot be assessed by our datasets. On the other
hand, HIS absorption features associated with strong stellar winds produced
by high star formation activity capable of ionizing radiation in high stellar
mass galaxies with high star formation activity sweeping  material to the
outer regions of these galaxies \citep
{Putman17,Bower16,Oppenheimer16,Voit15b}. 

Similar analyses including observations at different redshifts coming from
different complementary surveys, for example  VANDELS \citep
{McLure18,Pentericci18}, zCOSMOS \citep{Lilly07,Lilly09}, VVDS \citep
{LeFevre13b}, and  DEIMOS10K \citep{Hasinger18}, can help to increase the
number of galaxy pairs with close angular separations ($b<6\arcsec$), deblend
close spectral line features(e.g. C \,{\sc ii}--O \,{\sc iv}
$\uplambda\uplambda$ 1334.5, 1343.0 $\AA$; Si \,{\sc iv} $\uplambda\uplambda$
1393.8, 1402.8 $\AA$), and cover different low- and high-ionization state
lines(e.g.  O \,{\sc vi} $\uplambda\uplambda$ 1032, 1038 $\AA$,  Mg \,
{\sc ii} $\uplambda$ 2798 $\AA$). Additionally, it has been shown that  AGN
activity is dependent on stellar mass and SFR (e.g. \citealt
{Lemaux14,Bongiorno16,Magliocchetti20}), and thus studies considering \textit
{bg}--AGN and/or QSO--AGN galaxy pairs \citep{Hennawi06,Prochaska14} are
needed to study the effect that the presence of an AGN has on feedback and
quenching in star-forming galaxies at low and high redshifts.

\subsection{C\,{\sc ii} / C\,{\sc iv} line ratio.} 

A further inspection of the differences between LIS and HIS on the different
galaxy parameters explored comes from the  C\,{\sc ii}/C\,{\sc iv} line
ratio. We find that C\,{\sc ii}/C\,{\sc iv} line ratio anticorrelates with
impact parameter, stellar mass, and star formation rate.  This implies that
C\,{\sc iv} is mostly located at larger distances in more massive galaxies
with higher star formation rates, while C\,{\sc ii} dominates at smaller
impact parameters and is found mostly in low stellar mass galaxies with low
star formation rates. These results further support the picture of a
multi-phase CGM where LIS line absorptions are produced by denser gas with
lower temperatures (T$<10^{4.5}$K) that is located close to the central
galaxy, while HIS line absorptions are produced by warm gas located at larger
distances from the central galaxy. Simulations have shown that much of the
ongoing gas accretion occurs towards the edges of the galaxies to avoid the
dominant feedback (outflows) from the central regions
\citep{Stewart12,Fernandez12,Putman17}; however, we do not find supporting
evidence of the low-redshift scenario where low-ionization state gas infalls
along the major axis of star-forming galaxies, accompanied by large-scale
outflows along the minor axis forming bipolar outflows. On the contrary, our
results suggest that this scenario is not broadly applicable at z$\sim$2. We
note that we have also explored Si\,{\sc ii} / Si\,{\sc iv} and  Al\,
{\sc ii} / Al\,{\sc iii} equivalent width line ratios; however, we did not
detect any significant correlations. On the one hand, Si\,{\sc iv}
predominantly arises in denser gas closer to galaxies more similar to other
LIS state lines such as Mg\,{\sc ii} than to those of higher ions \citep
{Ford13,Ford14}, and hence Si\,{\sc ii} and Si\,{\sc iv} line absorption
might be probing gas with similar physical conditions. On the other hand,
Al\,{\sc iii} is an intermediate-ionization state (IIS) tracer of moderately
photoionized warm gas generally associated with neutral phase gas \citep
{Savage01,Knauth03,Vladilo01}. In fact, Al\,{\sc iii} absorption is
associated with low ions, as shown by the significant correlation between
the velocity widths of Al\,{\sc iii} and low-ionization species (e.g. Si\,
{\sc ii}, Fe\,{\sc ii}, Zn\,{\sc ii}, \citealt{Howk99b}), suggesting that a
substantial fraction of the low-ionization ions may be associated with
moderately ionized gas traced by Al\,{\sc iii} and Fe\,{\sc iii}
\citep{Wolfe05} and that doubly ionized species (e.g. Al\,{\sc iii}, Fe\,
{\sc iii}, C\,{\sc iii}, Si\,{\sc iii}) often have comparable column
densities to the singly ionized species (e.g. Al\,{\sc ii}, Fe\,{\sc ii},
C\,{\sc ii}, Si\,{\sc ii})  \citep{Mas-Ribas17b}.

\subsection{Ly$\upalpha$ emission}

Regarding Ly$\upalpha_{\rm{em}}$, several studies have been carried out to
explore the relation with galaxy properties. At low redshift \cite
{Runnholm20} used the Lyman Alpha Reference Sample (LARS) to obtain
correlations between Ly$\alpha$ and different galaxy properties (i.e. star
formation rate, dust extinction, compactness, and the gas covering fraction).
At high redshift it has been found that bluer galaxies show stronger
Ly$\upalpha_{\rm{em}}$
\citep{Shapley03,Pentericci10,Berry12,Erb16}, and that low stellar mass
galaxies with lower SFRs show stronger Ly$\upalpha_{\rm{em}}$ than
high-stellar mass galaxies with higher SFRs \citep
{Vanzella09,Stark10,Erb06a,Jones12,Shapley03,Kornei10,Hathi16}. Additionally
Ly$\upalpha_{\rm{em}}$ is correlated with LIS absorption; stronger LIS
absorptions correspond to weaker Ly$\upalpha_{\rm{em}}$, as high
concentrations of neutral gas responsible for strong LIS absorption are also
responsible for scattering out of the l.o.s. Ly$\upalpha$ photons \citep
{Shapley03,Vanzella09,Jones12}. More recently, \citet{Du18},
\citet{Trainor19}, and \citet{Pahl20} studied the spectroscopic properties of
star-forming galaxies at $z\sim$2--5 through composite spectra grouped
through different galaxy properties. They detected stronger Ly$\upalpha_{\rm
{em}}$ at higher redshift at fixed stellar mass, SFR, and UV luminosity, and
found that the LIS--Ly$\upalpha_{\rm{em}}$ relation is redshift independent,
suggesting that this is caused by the variations of the neutral gas covering
fraction favouring Ly$\alphaup$ escape and production and/or dust content in
the ISM and CGM. Moreover \cite{Oyarzun16,Oyarzun17} show that at high
redshift ($3<z<4.6$) the Ly$\upalpha_{\rm{em}}$ anticorrelations with
stellar mass, star formation rate, and UV luminosity are stronger in low
stellar mass populations, which is  explained by the rapidly increasing
neutral gas fraction of the universe at higher redshifts \citep{Du21}.



\section{Conclusions}\label{sec:Con}

In this paper we presented stacks of 238 \textit{background} galaxy spectra used to probe the
CGM content and extent around star-forming galaxies at $\langle
z \rangle \sim 2.6$. We only used  spectra with highly reliable spectroscopic
redshifts ($95-100\%$ probability  of being correct) to identify the
low-ionization state (LIS: Si\,{\sc ii}, C\,{\sc ii}) and high-ionization
state (HIS: C\,{\sc iv}, Si\, {\sc iv}) metal absorption lines, and constrain
their spatial distribution and their dependence on stellar mass, star
formation rate, azimuthal angle, and effective radius.

We summarize our main results below:
   \begin{enumerate}
      \item We detect LIS and HIS metal absorption lines in the CGM around
       star-forming galaxies at distances up to 172\,kpc  and 146\,kpc, respectively.
       The limitations owing to the size of our sample did not
       allow us to follow metal lines beyond these distances. The strength of
       these absorptions decreases at increasing distances from the galaxy,
       consistently with results published in the literature. At any fixed
       distance from the galaxy, the strength of all absorption lines that we
       identify in our sample at $z \sim 2.6$ is greater than any other
       measurement at lower redshift, providing evidence of a redshift
       evolution of the CGM gas content responsible for producing these
       absorptions.
      \item We do not find any significant correlation between the LIS--HIS
       absorptions and the azimuthal angle ($\phi$). This is opposed to the
       scenario at low redshift where cold gas (traced by LIS line) is
       infalling onto galaxies along the plane containing the disc, while the
       gas heated and processed by star formation (traced by HIS lines) is
       outflowing perpendicularly to the plane. This can be explained by the
       fact that high-redshift galaxies have not formed a stable disc capable
       of collimating galactic winds into bipolar outflows. However, due to
       the small sample of close pairs with available morphological features,
       these trends need to be confirmed by applying a similar analysis to
       larger samples at these high redshifts.
      \item We find an anticorrelation between Ly$\upalpha_{\rm{em}}$ and the
       impact parameter $b$, in agreement with previous results at
       high redshift. 
      \item To assess the relative importance of LIS and HIS absorptions, we
       computed the C{\sc ii}/C{\sc iv} equivalent width line ratio and find
       that it correlates with impact parameter $b$, stellar mass, and star
       formation rate. The C{\sc ii}/C{\sc iv} $W_{0}$ line ratio is higher at
       small separations, mainly detected in star-forming galaxies
       ($\langle z \rangle \sim 2.6$) with low stellar masses and low star
       formation rates. Conversely, low C{\sc ii}/C{\sc iv} line ratios are
       defined by stronger C{\sc iv} line absorption compared to C{\sc ii}
       and are observed at large separations in star-forming galaxies with higher
       stellar masses and star formation rates.
   \end{enumerate}

The results presented here provide observational evidence consistent with a
scenario where star-forming galaxies at $\langle z \rangle \sim 2.6$ possess
a multi-phase CGM where LIS metal absorptions are produced by denser gas,
which is more extended in these star-forming galaxies. Our results suggest
that galaxies with higher star formation rates and high stellar masses have
stronger ionizing fluxes that are  able to ionize gas at larger distances
and/or are capable to sweep out the highly ionized gas (traced by C\,
{\sc iv}) farther away from the galaxy compared with less massive and less
star-forming galaxies. Subsequently, star-forming galaxies with low SFR and
low stellar mass show larger reservoirs of cold gas as probed by their C\,
{\sc ii} and  C\,{\sc iv} lines and their C\,{\sc ii}/C\,{\sc iv} equivalent
width line ratio. These large reservoirs of cold gas could be funnelled into
the galaxies and eventually provide the necessary fuel to sustain star
formation activity. Recently, \cite{WangS22} has demonstrated that
large-scale environment modulates star formation by regulating the way in
which galaxies breathe material in and out (accrete and expell)  by
exchanging material within the CGM. This process is synchronized with star
formation rate events occurring within a galaxy, and is related to maxima
(minima) of SFR associated with a previous decrease(increase) in the cold
circumgalactic gas phase, that halts a further rise(decline), and leads to a
fall(rise) in the star formation rate at later stages. Our results from C
{\sc ii} peaks and C{\sc iv} troughs detected in low-mass galaxies (low SFRs)
and high-mass galaxies (high SFRs) could be interpreted as the snapshots of
these two different stages of the complex interplay between the ISM, CGM, and
IGM in which galaxies exchange material. However, we note that
high-resolution observations are required to detect outflows expelling
material or inflows accreting cold material back to the galaxy as recycled
material. We highlight that although stacking increases the S/N of
the \textit{background} \textit
\emph{{spotlights}}, allowing us  to detect the faint signal produced by gas
in the CGM, it also smears out the information about the spatial, kinematic,
and ionization properties of the CGM. Thus, we reiterate the necessity to
perform deeper observations with higher resolutions at this and higher
redshift of multiple LIS and HIS metal lines to provide better constraints
on the properties of the multi-phase CGM. It is very likely that future and
ongoing observations from large and deep spectroscopic surveys such as
VANDELS \citep{McLure18,Pentericci18}, zCOSMOS \citep
{Lilly07,Lilly09}, VVDS\citep{LeFevre13b} or DEIMOS10K \citep
{Hasinger18} and Integral Field Unit observations from MUSE/KMOS/JWST will
help us understand more about the multi-phase nature of the CGM. Future work
should point towards a resolved view of the velocity field of the in-falling
and out-flowing gas of galaxies at redshift $z > 2$.



\begin{acknowledgements}
        We thank an anonymous referee for constructive comments and
        suggestions that improved the manuscript. Based on data obtained with
        the European Southern Observatory Very Large Telescope, Paranal,
        Chile, under Large Programs 175.A-0839, 177.A-0837, and 185.A-0791.
        This work is based on data products made available at the CESAM data
        center, Laboratoire d'Astrophysique de Marseille. This work partly
        uses observations obtained with MegaPrime/MegaCam, a joint project of
        CFHT and CEA/DAPNIA, at the Canada-France-Hawaii Telescope
        (CFHT) which is operated by the National Research Council (NRC) of
        Canada, the Institut National des Sciences de l'Univers of the Centre
        National de la Recherche Scientifique (CNRS) of France, and the
        University of Hawaii. This work is based in part on data products
        produced at TERAPIX and the Canadian Astronomy Data Centre as part of
        the Canada-France-Hawaii Telescope Legacy Survey, a collaborative
        project of NRC and CNRS. HMH acknowledge partial support from
        National Fund for Scientific and Technological Research of Chile
        (\emph{Fondecyt}) through grants no. 1171710 \& 1150216.
        E.I.\ acknowledges partial support from FONDECYT through grants
        N$^\circ$\,1221846 and 1171710.
        MA acknowledges support from FONDECYT grant
        1211951, "ANID+PCI+INSTITUTO MAX PLANCK DE ASTRONOMIA MPG 190030"
        and "ANID+PCI+REDES 190194". RA acknowledges support from ANID
        FONDECYT Regular Grant 1202007. We thank ESO staff for their support
        for the VUDS survey, particularly the Paranal staff conducting the
        observations and Marina Rejkuba and the ESO user support group in
        Garching. This research used Astropy,\footnote
        {http://www.astropy.org} a community-developed core Python package
        for Astronomy \citep{astropy:2013, astropy:2018}, Numpy \citep
        {numpy:2020}, Scipy \citep{SciPy:2020} and Matplotlib \citep
        {matplotlib:2007}
\end{acknowledgements}

\bibliographystyle{aa} 
\bibliography{References_2} 

\begin{appendix} \label{sec:Appendix} 
\section{Supplementary tables.} \label{sec:AppendixTbl}
\FloatBarrier
\begin{table}[h]
\centering
\caption{\label{tbl:PairsStat-STM}
\textit{Foreground--background} (\textit{fg-bg}) galaxy pair statistics 
according to the stellar mass (log[M$_{\bigstar}$/M$_{\odot}$]).}
\begin{tabular}{cccccc}
  \\
  \hline\hline
  \multicolumn{1}{c}{ID} &
  \multicolumn{1}{c}{N} &
  \multicolumn{1}{c}{$\langle$ log[M$_{\bigstar}$/M$_{\odot}$] $\rangle$} &
  \multicolumn{1}{c}{$\langle b \rangle$ (kpc)} & 
  \multicolumn{1}{c}{$\langle \rm{z}_{\rm{fg}}\rangle$} &
  \multicolumn{1}{c}{$\langle \rm{z}_{\rm{bg}}\rangle$} \\
  \multicolumn{1}{c}{c1} & 
  \multicolumn{1}{c}{c2} &
  \multicolumn{1}{c}{c3} &
  \multicolumn{1}{c}{c4} &
  \multicolumn{1}{c}{c5} &
  \multicolumn{1}{c}{c6} \\
  \hline\hline  
ALL & 238 &  9.73 $\pm$ 0.40 & 125.05 & 2.60 & 3.04\\
S1  &  61 &  9.26 $\pm$ 0.18 & 125.12 & 2.48 & 2.87 \\
S2  &  63 &  9.60 $\pm$ 0.07 & 119.63 & 2.65 & 3.10 \\
S3  &  65 &  9.87 $\pm$ 0.09 & 130.42 & 2.58 & 3.04 \\
S4  &  49 & 10.31 $\pm$ 0.24 & 124.83 & 2.71 & 3.19 \\
  \hline\hline
\end{tabular}
\tablefoot{
Column 1: Sample ID; 
Column 2: number of galaxies per sample; 
Column 3: mean stellar mass;
Column 4: mean impact parameter in kpc;
Column 5: mean redshift of the \textit{foreground} galaxies sample; and
Column 6: mean redshift of the \textit{background} galaxies sample.}
\end{table}

\begin{table}[h]
\centering
\caption{\label{tbl:PairsStat-SFR}
\textit{Foreground--background} (\textit{fg-bg}) galaxy pair statistics 
according to the star formation rate (log[(M$_{\odot}$yr$^{-1})]$).}
\begin{tabular}{cccccc}
  \\
  \hline\hline
  \multicolumn{1}{c}{ID} &
  \multicolumn{1}{c}{N} &
  \multicolumn{1}{c}{$\langle$ log[SFR/(M$_{\odot}$yr$^{-1})]$ $\rangle$ }&
  \multicolumn{1}{c}{$\langle b \rangle$ (kpc)} & 
  \multicolumn{1}{c}{$\langle \rm{z}_{\rm{fg}}\rangle$} &
  \multicolumn{1}{c}{$\langle \rm{z}_{\rm{bg}}\rangle$} \\
  \multicolumn{1}{c}{c1} & 
  \multicolumn{1}{c}{c2} &
  \multicolumn{1}{c}{c3} &
  \multicolumn{1}{c}{c4} &
  \multicolumn{1}{c}{c5} &
  \multicolumn{1}{c}{c6} \\
  \hline\hline  
ALL & 238 & 1.37 $\pm$  0.39 & 125.10 & 2.60 & 3.04 \\
S1  &  60 & 0.91 $\pm$  0.39 & 117.92 & 2.56 & 2.98 \\
S2  &  62 & 1.21 $\pm$  0.06 & 128.63 & 2.64 & 3.08 \\
S3  &  59 & 1.49 $\pm$  0.09 & 124.50 & 2.54 & 3.02 \\
S4  &  59 & 1.92 $\pm$  0.20 & 129.21 & 2.67 & 3.08 \\
  \hline\hline
\end{tabular}
\tablefoot{
Column 1: Sample ID; 
Column 2: number of galaxies per sample; 
Column 3: mean star formation rate (log[SFR/(M$_{\odot}$yr$^{-1})]$);
Column 4: mean impact parameter in kpc;
Column 5: mean redshift of the \textit{foreground} galaxies sample; and
Column 6: mean redshift of the \textit{background} galaxies sample.}
\end{table}

\begin{table}[h]
\centering
\caption{\label{tbl:PairsStat-PHI}
\textit{Foreground--background} (\textit{fg-bg}) galaxy pair statistics 
according to the azimuthal angle ($\phi$).}
\begin{tabular}{cccccc}
  \\
  \hline\hline
  \multicolumn{1}{c}{ID} &
  \multicolumn{1}{c}{N} &
  \multicolumn{1}{c}{$\langle$ $\phi \rangle$ ($^{\circ}$)}&
  \multicolumn{1}{c}{$\langle b \rangle$ (kpc)} & 
  \multicolumn{1}{c}{$\langle \rm{z}_{\rm{fg}}\rangle$} &
  \multicolumn{1}{c}{$\langle \rm{z}_{\rm{bg}}\rangle$} \\
  \multicolumn{1}{c}{c1} & 
  \multicolumn{1}{c}{c2} &
  \multicolumn{1}{c}{c3} &
  \multicolumn{1}{c}{c4} &
  \multicolumn{1}{c}{c5} &
  \multicolumn{1}{c}{c6} \\
  \hline\hline  
ALL &  97 & 31.34 $\pm$ 18.12 & 121.54 & 2.65 & 3.08 \\
S1  &  60 & 16.90 $\pm$  6.20 & 107.74 & 2.69 & 3.15 \\
S2  &  62 & 44.92 $\pm$  6.74 & 132.53 & 2.55 & 2.92 \\
S3  &  59 & 75.13 $\pm$  9.10 & 118.75 & 2.70 & 3.16 \\
  \hline\hline
\end{tabular}
\tablefoot{
Column 1: Sample ID; 
Column 2: number of galaxies per sample; 
Column 3: mean azimuthal angle ($\phi \, ^{\circ}$);
Column 4: mean impact parameter in kpc;
Column 5: mean redshift of the \textit{foreground} galaxies sample; and
Column 6: mean redshift of the \textit{background} galaxies sample.}
\end{table}

\begin{table}[h]
\centering
\caption{\label{tbl:PairsStat-REFF}
\textit{Foreground--background} (\textit{fg-bg}) galaxy pair statistics 
according to the effective radius (r$_{\rm{eff}}$ \, kpc).}
\begin{tabular}{cccccc}
  \\
  \hline\hline
  \multicolumn{1}{c}{ID} &
  \multicolumn{1}{c}{N} &
  \multicolumn{1}{c}{$\langle$ r$_{\rm{eff}}$ $\rangle$ (kpc) }&
  \multicolumn{1}{c}{$\langle b \rangle$ (kpc)} & 
  \multicolumn{1}{c}{$\langle \rm{z}_{\rm{fg}}\rangle$} &
  \multicolumn{1}{c}{$\langle \rm{z}_{\rm{bg}}\rangle$} \\
  \multicolumn{1}{c}{c1} & 
  \multicolumn{1}{c}{c2} &
  \multicolumn{1}{c}{c3} &
  \multicolumn{1}{c}{c4} &
  \multicolumn{1}{c}{c5} &
  \multicolumn{1}{c}{c6} \\
  \hline\hline  
ALL &  97 &  8.94 $\pm$ 0.49 & 121.54 & 2.65 & 3.08 \\
S1  &  31 &  2.70 $\pm$ 0.98 & 125.92 & 2.67 & 3.16 \\
S2  &  33 &  6.87 $\pm$ 1.40 & 120.75 & 2.63 & 3.02 \\
S3  &  33 & 16.88 $\pm$ 7.39 & 118.19 & 2.66 & 3.07 \\
  \hline\hline
\end{tabular}
\tablefoot{
Column 1: Sample ID; 
Column 2: number of galaxies per sample; 
Column 3: mean effective radius (r$_{\rm{eff}}$ \, kpc);
Column 4: mean impact parameter in kpc;
Column 5: mean redshift of the \textit{foreground} galaxies sample; and
Column 6: mean redshift of the \textit{background} galaxies sample.}
\end{table}

\end{appendix}
\end{document}